\documentclass[journal=rsm]{CUP-JNL-DTM}%

\usepackage{graphicx}
\usepackage{multicol,multirow}
\usepackage{amsmath,amssymb,amsfonts}
\usepackage{mathrsfs}
\usepackage{amsthm}
\usepackage{rotating}
\usepackage{appendix}
\usepackage{ifpdf}
\usepackage[T1]{fontenc}
\usepackage{newtxtext}
\usepackage{newtxmath}
\usepackage{textcomp}
\usepackage{xcolor}
\usepackage{lipsum}
\usepackage[colorlinks,allcolors=blue]{hyperref}
\usepackage{booktabs}

\theoremstyle{definition}

\numberwithin{equation}{section}

\jname{Research Synthesis Methods}
\articletype{RESEARCH ARTICLE}
\jyear{2025}
 
\begin{document}

\providecommand{\expect}{\mathrm{E}}             
\providecommand{\var}{\mathrm{Var}}              
\providecommand{\normaldistn}{\mathrm{Normal}}   
\providecommand{\unifdistn}{\mathrm{Uniform}}    
\providecommand{\rmk}[1]{{[\textit{#1}]}}
\providecommand{\uisd}{\sigma_\mathrm{u}}
\providecommand{\Tau}{\mathrm{T}}
\providecommand{\expect}{\mathrm{E}}
\providecommand{\unifdistn}{\mathrm{Uniform}}

\begin{Frontmatter}

\title[Article Title]{Subgroup comparisons within and across studies in meta-analysis}

\author[1]{Renato Panaro}
\author[1]{Christian R\"{o}ver}
\author[1,2,3]{Tim Friede}

\authormark{Renato Panaro, Christian R\"{o}ver, Tim Friede}

\address[1]{\orgdiv{Department of Medical Statistics}, \orgname{University Medical Center G\"{o}ttingen}, \orgaddress{\state{G\"{o}ttingen}, \country{Germany}}.\email{rvpanaro@gmail.com}}
\address[2]{\orgname{DZHK (German Center for Cardiovascular Research)}, \orgdiv{partner site Lower Saxony}, \orgaddress{\state{G\"{o}ttingen}, \country{Germany}}}
\address[3]{\orgname{DZKJ (German Center for Child and Adolescent Health)}, \orgaddress{\state{G\"{o}ttingen}, \country{Germany}}}



\abstract{Meta-analyses of published subgroup-level data may yield different interaction
estimates depending on how the available data are weighted across studies. In
particular, the difference between meta-analytic subgroup-specific effects need
not equal the meta-analysis of within-study subgroup differences when these
quantities are based on different study weights. Variation in subgroup
prevalence across trials, including the presence of single-subgroup trials, may
further amplify this discrepancy and introduce aggregation bias into
across-study subgroup comparisons. We examine established approaches for
jointly or separately estimating subgroup-specific effects and
treatment-by-subgroup interactions and introduce the Same Weighting Across
Different Analyses (SWADA) principle. SWADA applies a common set of study
weights across the connected subgroup and interaction analyses, thereby
guaranteeing algebraic agreement between the difference of subgroup averages
and the average subgroup difference. Through simulation studies and
meta-analyses of randomised trials of COVID-19 treatments, we compare several
SWADA weighting schemes with existing approaches in terms of bias, coverage,
precision, and root mean squared error. SWADA ensures internally coherent
subgroup and interaction estimates, but may require down-weighting or excluding
studies that provide information for only one subgroup. The choice of weights
therefore involves a trade-off between algebraic coherence, use of the
available subgroup evidence, and statistical efficiency. SWADA should be viewed as a general weighting principle for meta-analytic comparisons rather than as a universally preferable estimator.}
\end{Frontmatter}



\section*{Highlights}
\paragraph{What is already known:}
\begin{itemize}
  \item Within- and across-trial interaction estimators often disagree
  \item Differences in subgroup prevalence across studies can amplify these discrepancies  
  \item If a trial contributes data for only one subgroup, it can influence the analysis and break algebraic consistency between estimates
\end{itemize}

\paragraph{What is new:}
\begin{itemize}
  \item Constant subgroup prevalence across studies (not necessarily balanced) guarantees agreement between within- and across-trial estimators
 
  \item Existing estimate-matching approaches may fail to provide reliable estimators when aggregation bias depends on subgroup composition

 \item We introduce SWADA, a weighting-based approach that guarantees agreement between subgroup-specific and interaction effect estimators

  \item Some SWADA variants may assign zero weight to single-subgroup trials, creating a trade-off between algebraic consistency and use of all available subgroup evidence
\end{itemize}

\paragraph{Potential impact for RSM readers outside the authors' field:}
\begin{itemize}
  \item The insights provided into interaction meta-analyses can greatly benefit healthcare policymakers and administrators by enabling more informed, evidence-based healthcare policies, particularly in resource allocation and the approval of new personalized therapies based on unbiased interaction and matching subgroup estimates
  \item Researchers from interdisciplinary fields, including health economics, public health, and epidemiology, can use the methodological statements (average difference vs. difference in averages) presented in this paper
\end{itemize}

\section{Introduction}\label{sec:intro}
Besides an \emph{overall} treatment effect (combined-population), clinical trials of interventions commonly also report results for certain subgroups of patients within the overall study population. Patient subgroups are distinguished based on (often binary or dichotomized) patient-level characteristics (\emph{moderators}). Clinical trials are usually designed for investigating the overall treatment effect, while subgroup investigations (Do subgroups benefit equally? Do all of them benefit? Are some harmed?) often remain inconclusive due to insufficient power. Nevertheless, they are frequently reported with the aim of demonstrating homogeneity of treatment effects across subgroups, although in some cases they may reveal harm in specific subgroups, suggesting that treatment should be withheld or stopped. Technically speaking, the exploration of subgroups means investigating a (``treatment-by-subgroup'') \emph{interaction effect}.
An \emph{interaction meta-analysis} serves to amplify the power of looking into interactions by aggregating data from multiple trials investigating the same clinical intervention [\cite{daCosta2019}]. 

Interaction meta-analyses have attracted attention in the context of the meta-analyses of COVID-19 trials [\cite{Sterne2020, Selvaraj2022, Albuquerque2022}].
In this setting, discrepancies became apparent between the pooled interaction effects (the average within-study differences between patient subgroups) and the difference between the pooled subgroup-specific average effects (the difference in averages). Such contradictions, however, are not entirely unexpected. The difference in average (DA) effects  across studies and the average difference (AD) in (within-study) effects  do not necessarily need to be the same. In the present case, the phenomenon may be amplified by the fact that study sizes and prevalence of subgroups across studies were particularly varied, to the extent that not all studies included patients from all subgroups. This issue is particularly relevant in the broader discussion of subgroup analyses in clinical research, where misleading or over-interpreted subgroup-specific effects have led to
inappropriate clinical decisions. As noted in previous research on treatment heterogeneity, subgroup differences observed within a single study do not always generalize across trials, especially when patient distributions are unbalanced [\cite{Oxman1992, Oxman2012}].

It has previously been pointed out that in circumstances like these different types of evidence come into play, namely, \emph{study-generated} evidence that relates to effects observed \emph{within} studies, as well as \emph{synthesis-generated} evidence that originates from comparisons \emph{across} studies [\cite{Cooper2009xx}]. When these types of evidence are mixed, the resulting estimate may be subject to ``aggregation bias'' (also known as the \emph{ecological fallacy}) [\cite{BakerEtAl2009, ThompsonHiggins2002, GeissbuehlerEtAl2021,Godolphin2022}]. 
To illustrate this issue, consider a meta-analysis of studies investigating the effects of corticosteroids on mortality in COVID-19 patients; besides the effect in the overall population, it was of interest to investigate whether the effect depended on participants' need for invasive ventilation [\cite{Sterne2020}].
In this interaction meta-analysis, there was a clear mismatch when comparing the aggregated effects in
the invasive and noninvasive patient subgroups with the average (log-)difference in effects that were observed
in each study. While both comparisons agreed in favouring the noninvasive group, the average of interaction effects derived within each study was roughly twice as large as the apparent subgroup-specific effect when comparing the average effects in the invasive and noninvasive subgroups. For details we refer to Section~\ref{sec:Examples}. Such results may first of all be counter-intuitive, but these also highlight the importance of careful modelling and interpretation when going beyond ``simple'' separate subgroup-specific meta-analysis\@.

We investigate and compare existing approaches for meta-analysis of subgroup-specific effects as well as treatment-by-subgroup interactions, including van Houwelingen's model [\cite{vanHouwelingenEtAl1993}] as well as the recently proposed within-trial framework by Godolphin \emph{et~al}  [\cite{Godolphin2022}]. The origins of the discrepancies can be traced to differing weighting schemes. This work not only addresses theoretical aspects but also demonstrates practical implications through real-world examples and comprehensive simulation studies. We (i) formalize \emph{Same Weighting Across Different Analyses (SWADA)} estimators to enforce consistency between DA and AD while preserving consistent subgroup means; (ii) show that interaction-weights SWADA provide within-trial interaction effects under our framework; and (iii) demonstrate that some SWADA variants attain near-nominal coverage in simulations, often with wider intervals or reduced efficiency. We therefore present SWADA as a principle for guaranteeing algebraic consistency. SWADA may involve down-weighting or even excluding certain studies; as a result, single-subgroup trials may be excluded, which creates a trade-off between consistency and efficiency.

The structure of the current work is outlined as follows: Section~\ref{sec:Examples} presents motivating examples. Section~\ref{sec:stats} delineates the estimators under investigation. Section~\ref{sec:sim} offers a comprehensive simulation study. Section~\ref{sec:example} presents the application results. The discussion belongs to Section~\ref{sec:dis}.

\section{Motivating examples}\label{sec:Examples}
\subsection{Effects of IL-6 antagonists on 28-day all-cause mortality in hospitalised COVID-19 patients}\label{sec:CortiExampleIntro}
\begin{figure}[!h]
\centering  \includegraphics[width = 0.9\linewidth]{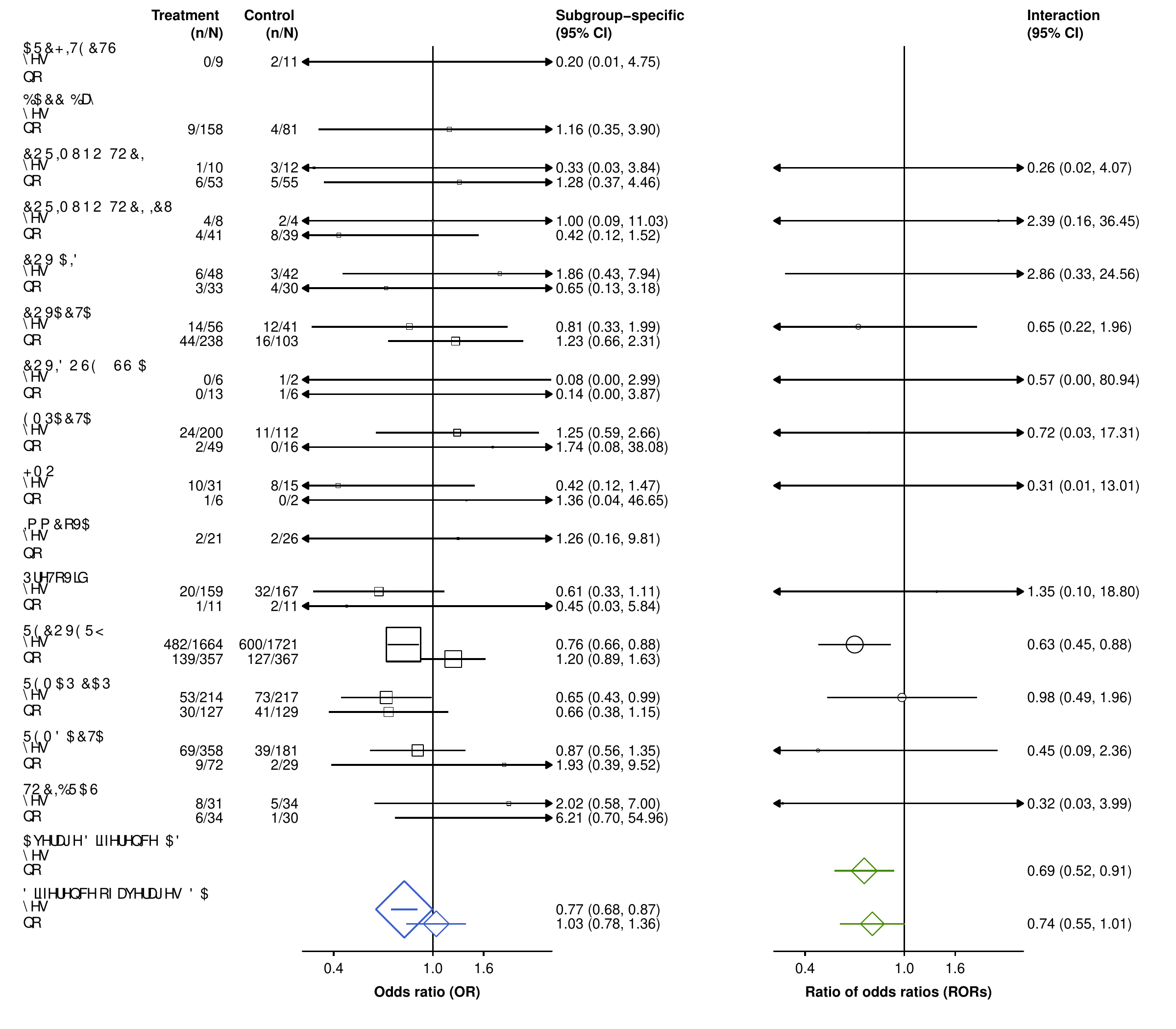}
\caption{Random-effects meta-analysis of the effect of IL-6 antagonists on mortality by corticosteroid administration (``yes'' vs. ``no'') in COVID-19 patients~[\cite{Shankar-Hari2021}]. The squares on the left represent the subgroup-specific treatment effects (ORs for patient subgroups with or without supplementing corticosteroid treatment). The bubbles on the right represent the  interactions (RORs for the subgroup difference within each study). Pooled effect (diamond) at the bottom refers to the (separate) meta-analyses of the above subgroup-specific effects or interactions. Here the (exponentiated) difference in average (DA) subgroup-effects (left-hand side) suggests a larger effect for patients \emph{with} corticosteroids by a factor of $0.77/1.03 = 0.74$ (rounded), while the (exponentiated) average of differences (AD) observed within studies (right-hand side) points to a factor of~$0.69$. Minor numerical differences between reported ORs and their corresponding RORs are due to rounding. Single-subgroup trials (ARCHITECTS, BACC-Bay, ImmCoVA) account for \(306\) of \(7,490\) participants, i.e. \(4.1\%\) of the total sample size in this meta-analysis}\label{fig:IL6_COVID19_Analysis}
 \end{figure}

In our first example, we use a meta-analysis of randomised clinical trials (RCTs) by the World Health Organization (WHO) Rapid Evidence Appraisal for COVID-19 Therapies (REACT) working group [\cite{Shankar-Hari2021}], also used by Godolphin \emph{et~al} (2023) [\cite{Godolphin2022}]; the dataset comprises data from 15~randomised clinical trials reporting on patients hospitalised for COVID-19. 
The meta-analysis aims to investigate the effects of interleukin-6 (IL-6) antagonist treatment (vs. control) on all-cause mortality, and a secondary focus was on whether treatment effects differed depending on whether or not patients underwent combined treatment with corticosteroids (data illustrated in Figure~\ref{fig:IL6_COVID19_Analysis}). Three types of IL-6 antagonists and two different dose levels were investigated.
Not all studies included both types of patients (with and without concomitant corticosteroid treatment), and possible clinical heterogeneity was anticipated in the analysis. Substantial variation in the prevalence of patient subgroups is noticeable, including a number of trials included only patients from one subgroup (using corticosteroids in all patients). Comparison of the average treatment effects between patients with and without corticosteroid therapy indicates a DA-ROR (log-ratio of odds ratios, ROR) of $0.77/1.03 \approx 0.74$ (rounded) using the random-effects model. Low between-trial heterogeneity was found in the steroids-no subgroup, more precisely \(\widehat{\tau}_{\mathrm{A}} \approx 0.025\) (steroids-no) and \(\widehat{\tau}_{B} \approx 0\) (steroids-yes). In contrast, the AD-ROR, which is derived \emph{within studies}, suggests a slightly more pronounced  interaction, with an ROR of~$0.69$. The between-study interaction heterogeneity (see Section~\ref{sec:UnivReMa}) was also found to be negligible ($\widehat{\tau}_\gamma = 0$) using the restricted maximum likelihood estimator (REML). 

\subsection{Effects of corticosteroids on 28-day all-cause mortality in hospitalised COVID-19 patients}\label{sec:ReactExampleIntro}
The second example discusses another prospective meta-analysis conducted by the WHO REACT working group, with a focus on the association between the administration of steroids and the induced reduction in 28-day all-cause mortality.
The meta-analysis incorporates data from patients treated with steroids and those in the control group, considering subgroups of individuals who were subjected to \emph{invasive mechanical ventilation (IV)} and those who were not \emph{(NIV)}~[\cite{Sterne2020}]. It is anticipated that the effect of corticosteroids might vary between patients receiving IV compared to those who do not require it, as, for example, the use of~IV is also associated with disease severity.
The meta-analysis involves data from seven RCTs; not all studies included both kinds of (IV and NIV) patients, so that 11~effect estimates (logarithmic odds ratios, log-ORs) are eventually combined. Throughout, treatment effects $y_{\mathrm{A}j}$ and $y_{\mathrm{B}j}$ and all derived
quantities (within-trial contrasts $g_j$, pooled effects, and
interactions) are expressed on the log-odds-ratio scale, so that
differences on this scale correspond to (log) ratios of odds ratios.
Results are exponentiated to odds ratios (ORs) and ratios of odds
ratios (RORs) for reporting in tables and figures.

\begin{figure}
  \centering
  \includegraphics[width = 0.9\linewidth]{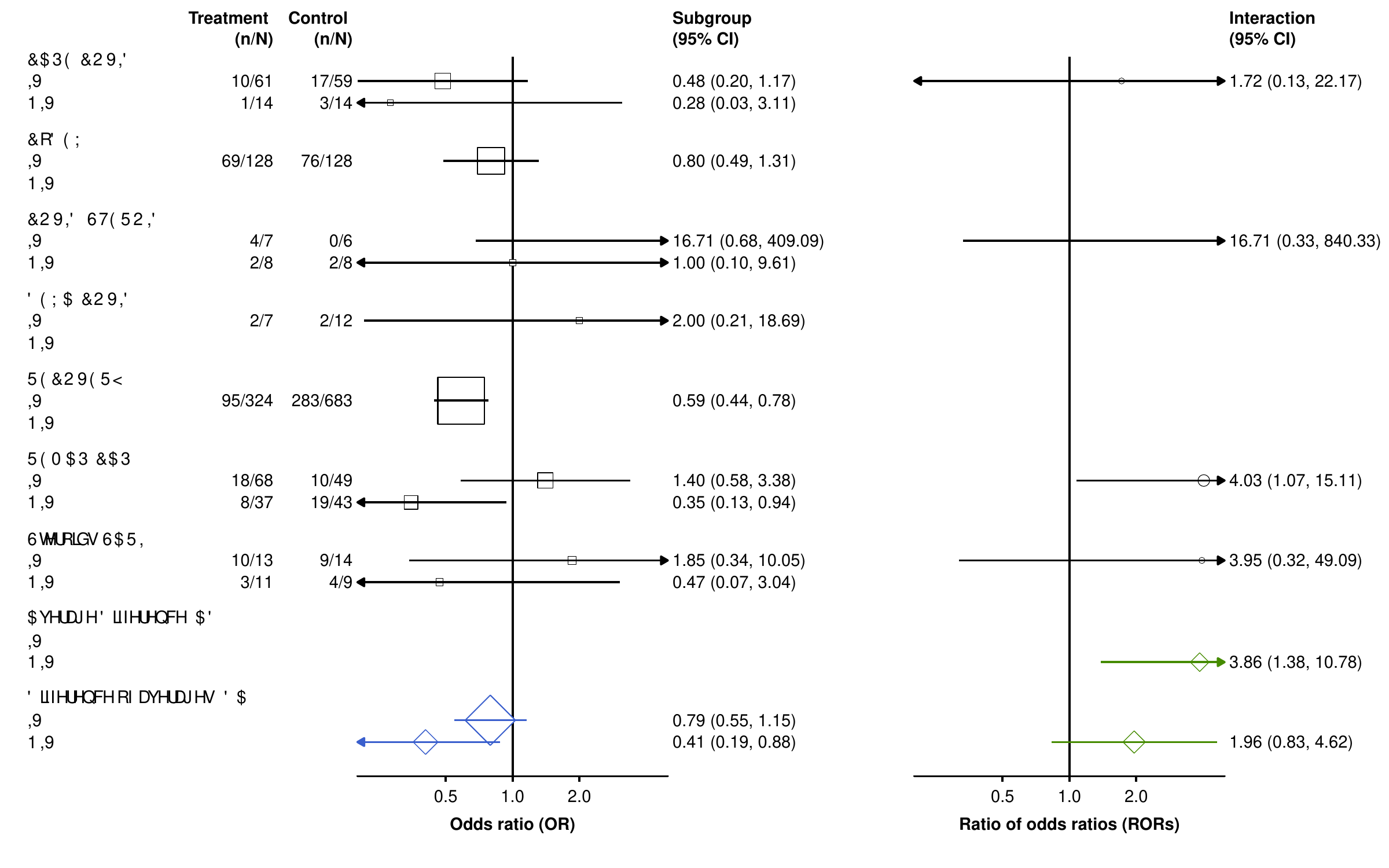}
  \caption{Random-effects meta-analysis of the effect of corticosteroid administration on mortality by invasive ventilation (``IV'' vs. ``NIV'') in COVID-19  patients~[\cite{Sterne2020}]. At the bottom, the meta-analysed pooled estimates (diamonds) are shown, highlighting the possibility of conflicting results: the mean IV and NIV~effects (ORs) differ by a factor of $0.79/0.41 \approx 1.96$ (rounded), i.e. the (exponentiated) difference of averages (DA) while the averaged interaction effect suggests a much larger ROR, or (exponentiated) average difference (AD), of~$3.86$. Minor numerical differences between reported ORs and their corresponding RORs are due to  rounding. Single-subgroup trials (CoDEX, DEXA-COVID~19, RECOVERY) account for \(1,282\) of \(1,703\) participants, i.e. \(75.3\%\) of the total sample size in this meta-analysis}\label{fig:Corticosteroids_MetaAnalysis}\centering \end{figure}

Figure~\ref{fig:Corticosteroids_MetaAnalysis} illustrates the (seemingly) discrepant meta-analysis~results when using different data aggregation schemes.
Computing the average treatment effect among IV~patients in all studies, and similarly among NIV~patients suggests a $1.68$-fold larger effect (OR) for IV patients under the CE (common-effect) assumption, and an approximate $2$-fold difference when incorporating heterogeneity, i.e. with estimated subgroup heterogeneities of \(\widehat{\tau}_{\mathrm{B}} \approx 0.075\) (IV) and \(\widehat{\tau}_{A} \approx 0\) (NIV) (see Figure \ref{fig:Corticosteroids_MetaAnalysis}). However, when computing ratios within each trial and aggregating these, one yields a much larger interaction (ratio of odds ratios, ROR) of~$\approx 3.9$ where  heterogeneity is estimated but found to be zero (\(  \widehat\tau_\gamma = 0)\).

\subsection{Dealing with single-subgroup trials}
\label{sec:Dealingwithsinglesubgrouptrials}
Including single-subgroup trials can contribute useful subgroup evidence, but when they carry large weight, as in the corticosteroid example (75.3\% of participants; Figure \ref{fig:Corticosteroids_MetaAnalysis}), subgroup analysis may not be appropriate regardless of method; their influence is much smaller in the IL-6 example (Figure 1). The COVID-19 examples include trials reporting results for only one corticosteroid subgroup (ARCHITECTS, BACC-Bay, and ImmCoVA in Figure~\ref{fig:IL6_COVID19_Analysis}) or for only one ventilation subgroup (CoDEX, DEXA-COVID 19, and RECOVERY in Figure~\ref{fig:Corticosteroids_MetaAnalysis}); that is, ``single-subgroup'' trials. In such cases, it is reasonable to question whether subgroup analysis is appropriate at all. The apparent discrepancy between the DA and the AD-RORs may simply reflect that these quantities are based on different data sets. Because information for the ``lacking'' subgroup in single-subgroup trials was represented as a null effect with an arbitrarily large variance in the subgroup analysis (left-hand side, OR-DA), but was entirely absent from the corresponding interaction analysis (right-hand side, AD-ROR; see Section~\ref{sec:notation}), the two approaches behave differently. Nevertheless, discrepancies of different magnitude persist even when these single-subgroup trials are excluded; in these examples, Appendix Figures \ref{fig:Final_Results_IL6_excluded} and \ref{fig:Final_Results_Corticosteroids_excluded} show DA-ROR = 0.74 vs. AD-ROR = 0.69 (IL6 antagonist-by-steroids) and AD-ROR = 3.9 vs. DA-ROR = 2.95 (Corticosteroid-by-ventilation), respectively. In both cases interaction weights, smaller subgroup and minimum of three weights SWADAs effectively down-weight or even exclude their influence (see Appendix \ref{app:Excluded}). We retain  single-subgroup trials in our re-analysis to keep consistency with originally published meta-analyses~[\cite{Sterne2020, Shankar-Hari2021}].

The non-inclusion of NIV endpoints from the Dexamethasone studies, for example, was justified in Sterne \emph{et~al} [\cite{Sterne2020}] due to the difficulty in assessing the critical illness status at the time of randomisation. The eventual impact of studies focusing on a single subgroup can be substantial, especially in large-scale studies. Incorporating a huge single-subgroup study predominantly influences the ``difference in means'' estimation, but does not affect the ``mean difference'', which may lead to seemingly incompatible results in interaction meta-analyses based on subgroup data.

In any single study, the subgroup and interaction estimates will always match; consider for example the first (``CAPE COVID'') study in Figure~\ref{fig:Corticosteroids_MetaAnalysis}: the ORs of~$0.48$ and~$0.28$ observed in the IV and NIV subgroups align with the interaction (ROR of~$0.48/0.28=1.72$) associated with this study. As we have already seen in the examples from Figures~\ref{fig:IL6_COVID19_Analysis} and~\ref{fig:Corticosteroids_MetaAnalysis}, as soon as we move to meta-analyses of subgroup-specific effects and interactions, the combined estimates do not necessarily match. Here we aim to shed some light on when and why this happens, and whether or how such seemingly contradictory results may be avoided or reconciled.
In the following, we introduce the terminology and notation, starting from the ``simple'' case of univariate meta-analysis, and then extend to data on patient subgroups within studies.

\section{Statistical methods and properties}\label{sec:stats}
\subsection{Notation for study and subgroup data} \label{sec:notation}
We consider the case of \emph{two} patient subgroups that are being contrasted in the meta-analysis of $k$~studies (as in the above examples).
The treatment effect for subgroup~$\mathrm{A}$ in the $j$th~trial is denoted by~${y}_{\mathrm{A}j}$, and the corresponding effect size for subgroup~$\mathrm{B}$ is~${y}_{\mathrm{B}j}$.
In matrix notation, ${y}_{\cdot j} = ({y}_{\mathrm{A}j}, {y}_{\mathrm{B}j})^{\prime}$ denotes the vector of both endpoints for the $j$th~study. 
Each of the two outcomes has a standard error~$\sigma_{ij}$ associated, resulting in a $2\times 2$ 
covariance matrix~$\Sigma_{j}$
for~$y_{\cdot j}$; the covariance matrix is diagonal, reflecting independence of subgroup estimates within trials because the subgroups do not overlap. Standard errors are, as usual, treated as fixed and known in the following, while in practice these are commonly estimated in an earlier stage of the analysis.
The $k$-dimensional row vector~$y_{i\cdot}$ denotes the outcomes for all~$k$ studies in their $i$th subgroup (with $i=\mathrm{A}$ or~$\mathrm{B}$).
Note that \emph{within} each study~$j$, we may derive an estimate of the  interaction, which results from the subgroup-specific effects as a scalar contrast $g_j = (y_{\mathrm{B}j}-y_{\mathrm{A}j})$ with standard error $\sqrt{\sigma_{\mathrm{A}j}^2+\sigma_{\mathrm{B}j}^2}$; the subgroup-specific effects and interactions are also shown in Figures~\ref{fig:IL6_COVID19_Analysis} and~\ref{fig:Corticosteroids_MetaAnalysis}.


Each study is based on a number~$n_j$ of participants, out of which $(1-p_j)\,n_j$ and $p_j\,n_j$ fall into subgroups~$\mathrm{A}$ and~$\mathrm{B}$, respectively; the~$p_{j}$ hence correspond to the \emph{subgroup B prevalences} in the $j$th study, that is, \emph{yes-steroids} and \emph{IV} in Figures~\ref{fig:IL6_COVID19_Analysis} and~\ref{fig:Corticosteroids_MetaAnalysis} respectively. Single-subgroup studies, discussed in Section~\ref{sec:Dealingwithsinglesubgrouptrials}, may also be accommodated in this terminology.
For example, a study only involving patients from subgroup~$\mathrm{A}$ (so that $p_j=0$) may be considered as providing an estimate ${y}_{\mathrm{B}j}$ with \emph{infinite} standard error~$\sigma_{\mathrm{B}j}$.
For any practical computation, the numerical value of~${y}_{\mathrm{B}j}$ then becomes irrelevant, and the associated contrast~$g_j$ likewise receives an associated infinite standard error. Many subsequent calculations (like the computation of ``inverse variance weights'') remain consistent (e.g., simply resulting in zero weights), however, in some cases it may also make sense to plug in some ``large'' number for the standard error (e.g., in the examples below, a
value of 100 was used), while inserting a ``neutral'' figure for~${y}_{\mathrm{B}j}$.
Following the convention in Godolphin \emph{et~al} [\cite{Godolphin2022}], we would then occasionally plug in~$g_j=0$ so that~${y}_{\mathrm{B}j} = {y}_{\mathrm{A}j}$.

\subsection{Approaches for subgroup meta-analysis }\label{sec:Methods}
\subsubsection{The univariate random-effects model}\label{sec:UnivReMa}
In the univariate case, each trial~$j$ provides a single estimate~$y_j$ of the treatment effect that has a standard error~$\sigma_j$ associated. It is assumed that for each trial estimates are (at least approximately) normally distributed around a true effect $\mu_j$:
\begin{eqnarray}
y_{j} \mid \sigma_j, \mu_j  \sim  \normaldistn\left(\mu_j, \sigma_j^2\right) 
\mbox{}\quad \mbox{and} \quad
\mu_j\mid \mu, \tau^2 \sim  \normaldistn\left(\mu,  \tau^2\right), \mbox{}
\end{eqnarray}
where ~$\tau$ represents the between-trial heterogeneity and $\mu$~is the pooled mean. This model is used to pool results across trials while accounting for both within- and across-trial variability.
In case $\tau=0$, the random-effects model reduces to the special case of a common-effect model (with $\mu_1=\ldots=\mu_k=\mu$) that was for example employed in the original analysis discussed in Section~\ref{sec:ReactExampleIntro}. The random-effects model is commonly used in practical applications and also was the basis for the three pooled estimates that were shown in Figures~\ref{fig:IL6_COVID19_Analysis} and~\ref{fig:Corticosteroids_MetaAnalysis}. The associated estimator of the pooled effect~$\mu$ eventually results as a weighted average of the (univariate)~$y_j$ [\cite{bayesmeta}].

This generic model may also be applied to subgroup-level data (subgroup-specific effects or interactions) as introduced in Section~\ref{sec:notation}; for example, for the interactions~$g_j$ one may then assume
\begin{equation}\label{eqn:interaction}
g_j  \mid \sigma_{\mathrm{A}j}, \sigma_{\mathrm{B}j}, \gamma_j \; \sim \; \normaldistn\left(\gamma_j, \sigma_{\mathrm{A}j}^2 + \sigma_{\mathrm{B}j}^2\right) 
\quad \mbox{and} \quad
\gamma_j\mid \gamma, \tau_\gamma^2 \; \sim \; \normaldistn\left(\gamma,  \tau_\gamma^2\right). \mbox{}
\end{equation}
We adopt the random-effects assumption throughout this work (\(\tau_\gamma \geq 0\)) and use the REML as the estimation method, as implemented in the \texttt{metafor} package in \textsf{R} [\cite{HedgesVevea1998,Viechtbauer2010}]. 

\subsubsection{Separate and joint pooling of (main) effects and treatment-by-subgroup interactions}\label{sec:separateAnalyses}
Pooling subgroup-specific effects and within-study contrasts separately involves shifting the focus from a single pooled efficacy measure, denoted by~$\mu$, to evaluating \emph{two} subgroup-specific effects in each study, denoted by the vector~$\beta$. This approach adds complexity to the analysis as it requires considering the correlation between subgroup random-effects. The examples shown in Section~\ref{sec:Examples} illustrate the practical implications of separate pooling, emphasizing the discrepancies that may arise.

Initially, in the previous Section~\ref{sec:UnivReMa}, we assumed a pooled mean model that could repeatedly be applied to pool subgroup-effects or interactions. This is in fact perfectly reasonable as long as analyses are viewed \emph{in isolation}. However, as soon as subgroups are contrasted, correlations in between-trial heterogeneity need to be considered. This necessitates the introduction of a between-trial heterogeneity matrix~($\Tau$) to describe the covariance structure of subgroup-specific effects. 
The following subsections will introduce several approaches to joint estimation based on the multivariate subgroup data.

\subsubsection{The bivariate model by van~Houwelingen, Arends, and Stijnen}\label{sec:vhas}
The model-based approach for subgroup-specific meta-analyses proposed by van~Houwelingen \emph{et~al} (2002) focuses on understanding how different non-overlapping subgroups of participants respond to treatments across multiple studies [\cite{vanHouwelingenEtAl2002}]. The approach combines two key components: the within-trial component, which looks at the treatment effect within each individual trial by considering specific participant characteristics (within-trial covariates), and the across-trial component, which examines variations in treatment effects between different trials. Using the terminology from Section~\ref{sec:notation}, the model is expressed in matrix notation as
\begin{equation}
y_{\cdot j} \mid  \Sigma_j, \beta_j \; \sim \; \normaldistn\left(\beta_j, \Sigma_j \right)
\quad \mbox{and} \quad
\beta_j \mid \beta, \Tau \; \sim \; \normaldistn\left(\beta, \Tau\right)\mbox{}
\end{equation}
After integrating over the study-specific effects, the resulting likelihood is given by
\begin{eqnarray}\label{eq:vhas}
\begin{pmatrix}{y}_{\mathrm{A}j} \\ {y}_{\mathrm{B}j}\end{pmatrix}  \mid \Sigma_j, \varphi, \gamma,  \Tau  & \sim & \normaldistn\left(\begin{pmatrix}
    \varphi \\ \varphi  + \gamma
\end{pmatrix}, \Sigma_j + \Tau \right)\mbox{,}
\end{eqnarray}
where  $\beta := (\beta_\mathrm{A}, \beta_\mathrm{B})^{\prime} = (\varphi, \varphi + \gamma)^{\prime}$ is a column-vector of constant coefficients in which $\varphi$ is the treatment effect on the reference subgroup and $\gamma$ denotes the difference between subgroup averages (\(\gamma = \beta_\mathrm{B}-\beta_\mathrm{A}\)).
The across-trial variability is now accounted for via a heterogeneity matrix $\Tau$. In the separate analyses case of Figures~\ref{fig:IL6_COVID19_Analysis} and~\ref{fig:Corticosteroids_MetaAnalysis} it is assumed to be diagonal. Later, the distinction is made between DA  (separate analyses using the univariate model from Section  \ref{sec:UnivReMa}) and the joint van Houwelingen's model (with an unstructured covariance matrix), as illustrated in Figures~\ref{fig:Final_Results_IL6} and~\ref{fig:Final_Results_Corticosteroids}.

This approach is essentially a bivariate generalization of the model from  Section~\ref{sec:UnivReMa}. By employing a joint model, a detailed and accurate understanding of subgroup-specific treatment effects across multiple clinical studies is achieved. Estimation within this model framework may be facilitated using generic software for mixed-effects models. The within-trial variance-covariance matrix~$\Sigma_j$ is usually diagonal, reflecting independence of subgroup estimates within trials. The between-trial heterogeneity~$\Tau$, on the other hand, would usually imply positive correlation between outcomes from the same study. Despite the more sophisticated modelling, estimates may still turn out ``contradictory'' (as exemplified in Section~\ref{sec:Examples}) with respect to subgroups and contrasts. 

\subsubsection{The within-trial framework} \label{sec:within-trial}

Another dedicated method for subgroup-data interaction meta-analysis is the \emph{within-trial framework} recently proposed by Godolphin \emph{et~al} [\cite{Godolphin2022}] (referred to as the ``within-trial framework'' in Table \ref{tab:sim-methods} and the figures). Motivated by the potentially contradictory results from subgroup-specific effect and interaction meta-analyses (see Section~\ref{sec:separateAnalyses}), it proceeds by prioritizing analysis steps, deriving the interaction estimate first, and subsequently deriving subgroup-specific effect estimates after fixing the interaction estimate obtained in the first step. The statistical model is specified as follows,
\begin{eqnarray}\label{eqn:withintrialframework}
\begin{pmatrix}
 {y}_{\mathrm{A}j} \\
 {y}_{\mathrm{B}j}-\widehat{\gamma}_W
\end{pmatrix} \mid  \Sigma_j, \widehat{\gamma}_W, \varphi, \Tau \;\sim\; \normaldistn\left(\begin{pmatrix}\varphi \\ \varphi\end{pmatrix},
\Sigma_j  + \Tau\right)
\end{eqnarray}
where $\widehat{\gamma}_W$ is the interaction estimate obtained with the
univariate within-interaction model previously discussed in
Section~\ref{sec:UnivReMa}. 
It is worth noting that the \emph{model assumptions} are similar to the previous section (Section~\ref{sec:vhas}); but the
estimation procedure differs. See Godolphin \emph{et~al} (2023)~[\cite{Godolphin2022}]
for details on the estimation procedure and also the case of more than two subgroups. Clearly, the resulting estimator above will
minimize the residuals of~\eqref{eqn:withintrialframework} when homogeneity
is assumed, but at the same time, it does not necessarily minimize the
residuals of~\eqref{eq:vhas}.

In its general case, the overall mean estimator corresponds to a precision-weighted average across studies, as obtained under REML estimation,
\(
\widehat{\varphi}
=
\frac{
\sum_{j=1}^k 
J^\top (\Sigma_j + \mathrm{T})^{-1}
(
y_{\mathrm{A} j},
y_{\mathrm{B} j} - \widehat{\gamma}_W
)^{\prime}
}{
\sum_{j=1}^k 
J^\top (\Sigma_j + \mathrm{T})^{-1} J
}\), where \(J\) is a column vector of ones. This expression simplifies, assuming disjoint subgroups (zero correlation) and $\mathrm{T} = 0$, to
\(
\widehat{\varphi}
=
\frac{
\sum_{j=1}^k 
\left(
\sigma_{\mathrm{A} j}^{-2} y_{\mathrm{A} j}
+
\sigma_{\mathrm{B} j}^{-2} (y_{\mathrm{B} j} - \widehat{\gamma}_W)
\right)
}{
\sum_{j=1}^k 
(\sigma_{\mathrm{A} j}^{-2} + \sigma_{\mathrm{B} j}^{-2})
}.
\)
The pooled subgroup-specific effects are then recovered as
\(\widehat{\beta}_{\mathrm{A}} = \widehat{\varphi}\), and
\(
\widehat{\beta}_{\mathrm{B}} = \widehat{\varphi} + \widehat{\gamma}_W.\)

Hence, each subgroup mean is a precision-weighted average of the corresponding study-specific estimates. The two-stage approach ensures consistency with the (fixed/known) univariate interaction estimate (as in Section~\ref{sec:UnivReMa}) as well as with the resulting subgroup means, at the cost of suboptimality elsewhere.

\subsubsection{The prevalence-adjusted model}\label{sec:WithinBetween}

A closely related context where data at an even greater level of detail may be used to investigate
interaction effects is the case of \emph{individual participant data (IPD)} meta-analysis.
Riley \emph{et~al} (2020)  underscore the importance of focusing on interaction effects \emph{within} studies in order to avoid the conflation of ``\emph{within-study}'' and ``\emph{between-study}'' evidence~[\cite{RileyEtAl2020}]. The influence of subgroup prevalences on each study's average effect is reflected in the model, where the interaction contributes to each subgroup average according to its subgroup proportion. We use the aggregation bias concept from IPD analysis (Appendix \ref{app:TheIPDprevalenceadjustedmodelbyRileyetal}) directly to an aggregate-data model (using the notation from Section~\ref{sec:notation}). 

\begin{figure}[!h]
    \centering
    \includegraphics[width= 0.9\linewidth]{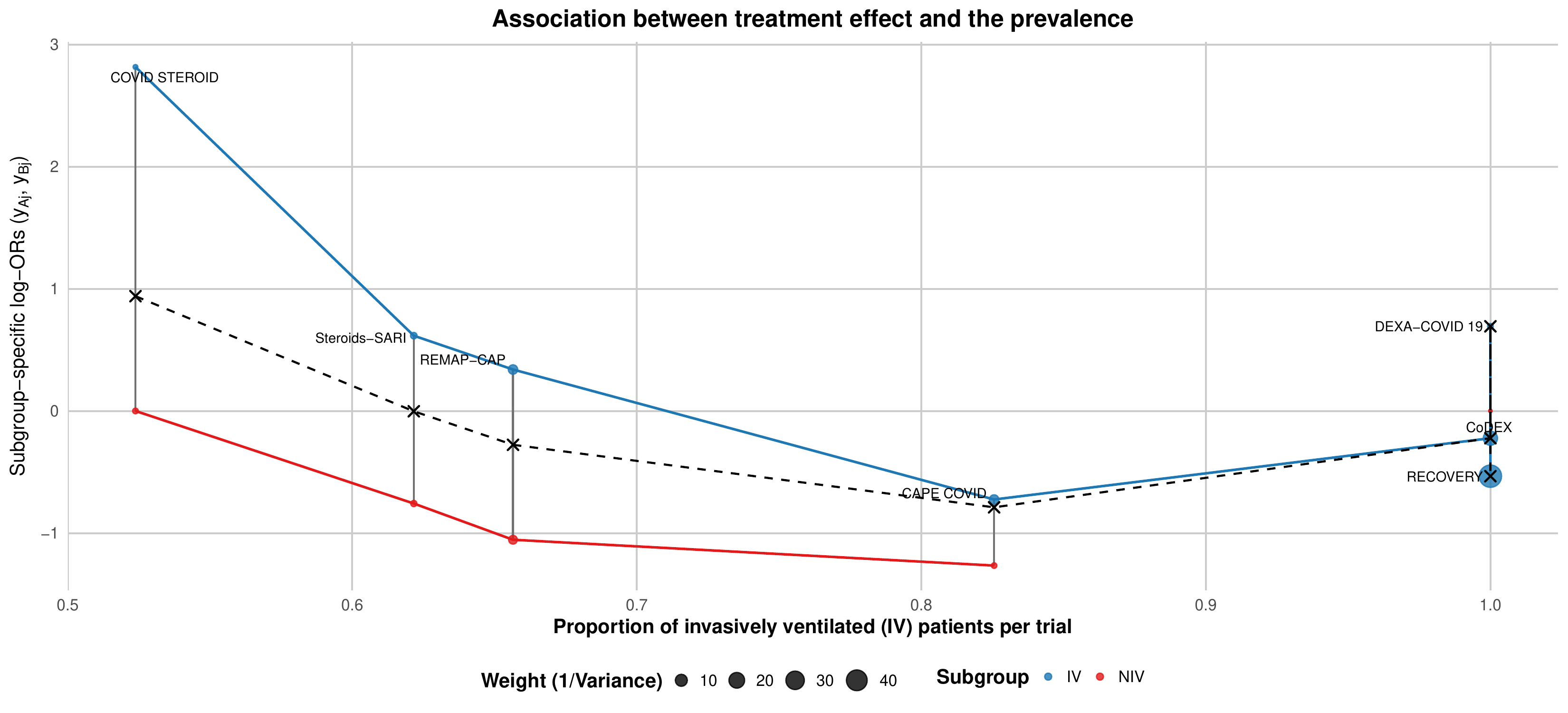}
    \caption{Subgroup-specific treatment effects (log-odds ratios) plotted against the prevalence of IV patients across trials. Points are colour-coded by subgroup (``IV'' vs ``NIV'') and sized by the inverse of their variance, reflecting relative precision. Grey lines connect the two subgroup estimates within each trial, illustrating the within-trial interaction (difference between IV and NIV effects). Apparent trends between subgroup-specific effects and prevalence can arise even without systematic aggregation bias, especially with few trials; the figure illustrates how such variation may appear in a few studies meta-analysis}    \label{fig:Subgroup_Prevalences}
\end{figure}

Figure \ref{fig:Subgroup_Prevalences} illustrates the variability of observed subgroup-specific effect estimates as a function of the seven studies' subgroup prevalences in the corticosteroids meta-analysis. First of all, only the point estimates are shown, and prevalences of IV~patients ranged from~55\% up to~100\% resulting in a less distinct appearance.
However, some of the main problems become evident here, in particular, the within-study evidence only stemming from four of the studies, and between-study evidence, and how it is also affected by single-subgroup studies that don't contribute to the within-study estimate.

We may account for the influence of subgroup prevalence on the average treatment effect~$\varphi$ by using a prevalence-outcome association model 
\begin{eqnarray}
\begin{pmatrix}{y}_{\mathrm{A}j} \\ {y}_{\mathrm{B}j}\end{pmatrix} \mid \Sigma_j, \varphi,  \gamma, \delta, p_j, \Tau_j & \sim & \normaldistn\left( 
\begin{pmatrix} \varphi + \delta p_j\\[1mm]
\varphi+ \delta p_j+\gamma \end{pmatrix},
\; \Sigma_j + \Tau_j \right). \label{eqn:PrevAdjModel}
\end{eqnarray}
where the between-study evidence for the interaction effect is denoted by ~\(\delta\). It is also worth considering the special case when all studies have the same, common subgroup prevalence~$p_j$ associated. In this case, the model simplifies;  subgroup proportions are not a source of additional variability, and subgroup-specific effects are directly comparable across trials. 

Although Section \ref{sec:Methods} presented these approaches as models (with their associated estimation procedures), each yields a point estimator that can be written explicitly as a weighted average of the input data. The estimators introduced above were mostly derived as maximum likelihood estimators based on particular modelling assumptions, or based on certain stepwise procedures. In all cases, it turned out that the eventual point estimators (for effects or interactions) resulted as \emph{weighted averages} of the input data (e.g., the study-specific contrast point estimates \(g_j\)). For a given quantity of interest (for example,~$\gamma$) the respective estimate is given by $\widehat{\gamma} = P_{\widehat \gamma} g_\cdot$, where $P_{\widehat \gamma}$ is the corresponding ``hat matrix'' defining a linear combination (projection) of the vector of individual interaction estimates ($g_\cdot$). 
It is important to note here that for the bivariate models, the projections are made \emph{jointly} using \emph{both} subgroup endpoints, that is, $(\widehat{\varphi}, \widehat{\varphi} + \widehat{\gamma})^{\prime} = P_{\widehat \beta} (y_{\mathrm{A}\cdot}, y_{\mathrm{B}\cdot})^{\prime}$. 
While the univariate analyses will always result in weighted averages (convex combinations) of the vector of point estimates, the multivariate case yields more flexible estimates, in particular, not necessarily convex weights, or even negative coefficients.

\subsection{Weighted estimators}\label{sec:WeightedAverages}

A general weighted average in the context of a univariate meta-analysis (see Section~\ref{sec:UnivReMa}) is given by $\widehat{\mu}=\sum_{j=1}^k w_j y_j$ (where $w_j \geq 0$ and $\sum_{j=1}^k w_j = 1$).
Commonly used estimates often employ the ``inverse variance'' weights $w_j=\frac{(\sigma_j^2+\widehat{\tau}^2)^{-1}}{\sum_{i=1}^k (\sigma_i^2 + \widehat{\tau}^2)^{-1}}$ that result both from the standard errors~$\sigma_i$ as well as the estimated heterogeneity~$\widehat{\tau}$ (that is,  \(\hat\tau_A\) for subgroup A,  \(\hat\tau_B\) for subgroup B or \(\hat\tau_\gamma\) for interaction).
Obvious alternative suggestions for the weights may include for example weighting by sample size ($w_j=\frac{n_j}{\sum_{i=1}^k n_i}$), which would coincide with the inverse-variance weights in case $\sigma_j=\frac{\uisd}{\sqrt{n_j}}$ and $\widehat{\tau}_\gamma=0$. Note that both \emph{sample size} and \emph{inverse variance} weighting schemes are also able to accommodate extreme (or limiting) cases of ``infinite standard errors'' (zero precision) or ``zero sample sizes'', in both cases simply leading to zero weight coefficients for certain studies.

The following section will shed more light on the connections between the internal workings of estimators and their potential to yield seemingly paradoxical results, which will in turn allow to devise estimators with certain desirable properties. Estimation via arbitrary weights is pretty simple and implemented e.g. in the \texttt{metafor} \textsf{R}~package  [\cite{HedgesVevea1998,Viechtbauer2010}].

\subsubsection{The Same Weighting Across Different Analyses (SWADA) principle}
\label{sec:CommonWeightEstimators}

The interaction (contrast of subgroup-specific effects) was defined as $\gamma=\beta_\mathrm{B}-\beta_\mathrm{A}$, however, we have already seen in the examples that this identity may not persist when it comes to the corresponding empirical estimates~$\widehat{\gamma}$, $\widehat{\beta}_\mathrm{A}$ and~$\widehat{\beta}_\mathrm{B}$ instead.
For this, consider the previously mentioned interaction estimator types.
\begin{itemize}
    \item The \emph{average subgroup difference} (AD) for any set of interaction weights \(w_{g\cdot}\), including interaction inverse-variance weights as a particular case (REML estimator in Section \ref{sec:UnivReMa}), is \begin{eqnarray}
\widehat{\gamma}_{\mathrm{AD}} &=& \sum\limits_{j = 1}^k w_{g_{j}} g_j =\sum\limits_{j = 1}^k w_{g_{j}} (y_{\mathrm{B}j} - y_{\mathrm{A}j}).\label{gamma2}
\end{eqnarray}
    \item The \emph{difference in subgroup averages} (DA) for any set of subgroup weights \((w_{{y_\mathrm{A}}_{\cdot}}, w_{{y_\mathrm{B}}_{\cdot}})^{\prime}\), including subgroup-specific inverse-variance weights as a particular case (REML estimator in Section \ref{sec:vhas}), is 
    \begin{equation}
\widehat{\gamma}_{\mathrm{DA}} \;=\; 
\sum\limits_{j = 1}^k w_{y_{\mathrm{B}j}}{y}_{\mathrm{B}j} - \sum\limits_{j = 1}^k w_{y_{\mathrm{A}j}}{y}_{\mathrm{A}j}\label{gamma1}.
\end{equation} 
Comparing~$\widehat{\gamma}_{\mathrm{DA}}$ and~$\widehat{\gamma}_{\mathrm{AD}}$, 
it becomes evident that both estimates are not necessarily identical, as we have already witnessed in the data examples above. 
However, one may also already spot cases where they will match, for example, if the study-specific weights are equal, that is, if $w_{g_j} = w_{y_{\mathrm{A}j}} = w_{y_{\mathrm{B}j}}$. We refer to any weighting based estimator satisfying the latter as a \emph{Same Weighting Across Different Analyses (SWADA)} estimator. We will examine these conditions more systematically in the following subsection and show that not only the estimates but also the quantities of interest may differ. 
\end{itemize}

Differing weighting schemes obviously come with different properties, in particular, with respect to bias or variance. 
Noting that we are free to define ``custom'' estimators for effects and interactions based on arbitrary pre-specified weights, we may then also be able to foster certain desirable properties [\cite{LairdMosteller1990}].
Such approaches have previously been utilized in other contexts. For example, Henmi and Copas (2011) noted in a similar context that any weighted average yielded an unbiased estimate, and that way they were able to specify estimators fixing certain problems resulting from correlations between effects and the commonly used inverse-variance weights [\cite{HenmiCopas2010}].

To get a feeling for the effects of certain weighting schemes, we may first consider the special case of ``inverse-variance'' weights (of the form $w_j=\frac{\sigma_j^{-2}}{\sum_i \sigma_i^{-2}}$). 
Starting with the simplest (usually rather optimistic) assumption of between-trial homogeneity~($\tau=0$), it is crucial to note that equal weightings (for subgroup averages as well as for the average subgroup contrast: $w_j = w_{g_j} = w_{y_{\mathrm{A}j}} = w_{y_{\mathrm{B}j}}$) will result whenever the ratio of subgroup variances
($\sigma_{\mathrm{A}j}^2/\sigma_{\mathrm{B}j}^2$) is the same for all studies~$j$. 
One may get an idea of when this condition should be met by considering the simple case when standard errors result as
$\sigma_{ij}=\frac{\uisd}{\sqrt{p_{ij}\,n_j}}$, where $\uisd$~is the same for both subgroups of a study [\cite{RoeverEtAl2021}]. 

In that case, the variance ratio simplifies to $\sigma_{\mathrm{B}j}^2/\sigma_{\mathrm{A}j}^2 = (1-p_j)/p_j$, meaning that constant \emph{subgroup prevalences} across studies will imply matching weights, and with that, matching contrast estimates.
Similar arguments hold if we consider the general case of arbitrary weights as the basis for the estimators~$\widehat{\gamma}_{\mathrm{DA}}$ and~$\widehat{\gamma}_{\mathrm{AD}}$; the crucial feature here is constancy of weight ratios for pairs of subgroups as a sufficient condition for agreement between the DA and AD estimators. 

\subsubsection{Properties of SWADA estimators}

SWADA defines a class of weighting-based estimators for subgroup-specific effects and treatment-by-subgroup interactions. It imposes a common study-weight structure across the connected analyses, so that the subgroup-specific pooled effects and the pooled within-study contrasts are constructed using the same study weights. This weighting constraint guarantees algebraic agreement between the difference of subgroup averages and the average subgroup difference. The particular choice of common weights determines the resulting estimand and its statistical efficiency.

The key idea is that the interaction can be estimated in two ways: by first pooling the subgroup-specific effects and then taking their difference, or by first calculating the subgroup difference within each study and then pooling these differences. For fixed common weights, under SWADA,
\(
w_{y_{\mathrm{A}j}}
=
w_{y_{\mathrm{B}j}}
=
w_{g_j}
=
w_j
\) for every \(j\)
and therefore
\begin{eqnarray}
\operatorname{E}\!\left(\widehat{\gamma}_{\mathrm{DA}}\right)
&=
\sum\limits_{j=1}^{k}
w_j\operatorname{E}\!\left(y_{\mathrm{B}j}\right)
-
\sum\limits_{j=1}^{k}
w_j\operatorname{E}\!\left(y_{\mathrm{A}j}\right) =
\sum\limits_{j=1}^{k}
w_j
\left[
\operatorname{E}\!\left(y_{\mathrm{B}j}\right)
-
\operatorname{E}\!\left(y_{\mathrm{A}j}\right)
\right] =
\operatorname{E}\!\left(\widehat{\gamma}_{\mathrm{AD}}\right).
\end{eqnarray}
Hence, for fixed weights, both estimators are unbiased for the same weighted interaction estimand,
\(
\sum_{j=1}^{k}
w_j\operatorname{E}\!\left(y_{\mathrm{B}j}-y_{\mathrm{A}j}\right)
\) regardless of the model assumed for the underlying data-generating process. For instance, if
\(
\operatorname{E}\!\left(y_{\mathrm{B}j}-y_{\mathrm{A}j}\right)=\gamma
\) for every \(j\)
and the weights sum to one, then
\(
\operatorname{E}\!\left(\widehat{\gamma}_{\mathrm{DA}}\right)
=
\operatorname{E}\!\left(\widehat{\gamma}_{\mathrm{AD}}\right)
=
\gamma.
\) Under SWADA, the DA and AD estimators are algebraically identical for a
given common weighting scheme and therefore have the same expected value
and variance. This equality does not imply that all SWADA weighting
schemes are equally efficient. The variance remains determined by the
chosen common weights. Assuming independence across trials, their variance under SWADA is
\begin{eqnarray}
\operatorname{Var}\!\left(\widehat{\gamma}_{\mathrm{DA}}\right)
=
\operatorname{Var}\!\left(\widehat{\gamma}_{\mathrm{AD}}\right)
=
\sum_{j=1}^{k} w_j^2
\operatorname{Var}\!\left(y_{\mathrm{B}j}-y_{\mathrm{A}j}\right).
\end{eqnarray}
This expression shows that the choice of \(w_j\), further discussed in
Section~\ref{sec:choosing}, primarily determines the variance of the SWADA
estimator and hence its efficiency. Interaction weights are designed to
minimize the variance of the interaction estimator, whereas alternative
weighting schemes may offer a different compromise between the precision of
the interaction and subgroup-specific estimators. In particular,
interaction-weights SWADA reproduces the conventional inverse-variance AD
estimator from \eqref{eqn:interaction} and its variance. Other SWADA weighting schemes may be less
efficient for estimating the interaction, but can provide a more balanced use
of uneven subgroup-specific information, as illustrated in
Section~\ref{sec:example}.

More generally, we define a quantity~\(\widehat{\Delta}(y)\) for generalised weighting projections that measures the discrepancy between these two routes. This quantity characterizes the mismatch between
\(\widehat{\gamma}_{\mathrm{DA}}\) and
\(\widehat{\gamma}_{\mathrm{AD}}\) (see the notation of Section~\ref{sec:WeightedAverages}). The following expressions represent this comparison in matrix form
\begin{eqnarray}\label{eqn:delta}
    \widehat{\Delta}({y}) &=& {\widehat{\gamma}_{\mathrm{DA}}}(y) - {\widehat{\gamma}_{\mathrm{AD}}}(y)\\  
      &=& C P_{\widehat \beta} \underbrace{(y_{\mathrm{A}\cdot}, y_{\mathrm{B}\cdot})^{\prime}}_{\text{subgroup-specific}} - P_{\widehat \gamma} \underbrace{g_\cdot}_{\text{interaction}} \\ \label{eqn:noncol}
      &=& C P_{\widehat \beta} (y_{\mathrm{A}\cdot}, y_{\mathrm{B}\cdot})^{\prime} - P_{\widehat \gamma} (C \otimes I_{k}) (y_{\mathrm{A}\cdot}, y_{\mathrm{B}\cdot})^{\prime} \\
      &=& D  (y_{\mathrm{A}\cdot}, y_{\mathrm{B}\cdot})^{\prime},
\end{eqnarray}
where \(C = [-1~1]\) is a contrast (row-) vector, \(\otimes\) denotes the Kronecker product, and \(P_{\widehat{\beta}}\) and \(P_{\widehat{\gamma}}\) are generic \(2 \times 2k\) and \(1 \times k\) projection matrices, respectively, with \(k\) denoting the number of trials. Here, \(g_\cdot = (C \otimes I_k )(y_{\mathrm{A}\cdot},y_{\mathrm{B}\cdot})^{\prime}\) represents the vectorized within-trial contrasts.

 Specifically, if the vectorized trial--subgroup-specific effect sizes \(y\) have mean \(\operatorname{E}(y)\) and variance--covariance matrix \(S\), the observed mismatch is approximately normally distributed as
\begin{eqnarray}\label{eqn:misma}
\widehat{\Delta}(y)
\sim
\operatorname{Normal}\!\left(D\operatorname{E}(y),DSD^{\prime}\right).
\end{eqnarray}
Here, \(D\operatorname{E}(y)\) represents the expected mismatch arising from different weighting projections, while \(DSD^{\prime}\) represents its variability. The  formulation distinguishes an expected agreement from algebraic
agreement. It is possible that
\(D\operatorname{E}(y)=0\) even when \(D\neq0\), in which case the DA and AD
estimators agree on average but may differ in an observed dataset, which translates to no aggregation bias in expectation. Their
anticipated discrepancy is then characterized by
\(DSD^{\prime}\). In contrast, under SWADA, \(D=0\), so that
\(\widehat{\Delta}(y)=0\) for every possible realization of the data, and
both its expected value and variance are zero. Additionally, the influence of
the \(j\)th study on the variability of this mismatch may then be assessed in a leave-one-out
analysis by comparing the variance--covariance matrix in~\eqref{eqn:misma} with and without that study. Possible summaries
include the ratio of variances [\cite{Viechtbauer2010o}] or, for multivariate
interactions, the determinant of the variance--covariance matrix.
Equivalently, one may maximize the determinant of the corresponding precision, as in the \(D\)-optimality criterion used in trial design.

\subsubsection{Choosing SWADA weights}\label{sec:choosing}

Motivated by the arguments made in the preceding Section~\ref{sec:WeightedAverages}, we may consider SWADAs that are applied to \emph{all three} types of meta-analyses (the two subgroup-specific effects and the interaction) in order to yield matching subgroup and interaction effect estimates. An important consequence is that enforcing this algebraic coherence may require down-weighting, or even assigning zero weight to, some studies. In particular, studies that contribute information to only one subgroup may be excluded under certain SWADA weighting choices, which can reduce the effective sample size.

Some candidates for SWADA weighting schemes may be
\begin{enumerate}
  \item \emph{Equal weights SWADA -- equal weights for all studies},  $w_j=\frac{1}{k}$:\\
  A somewhat ``obvious'' choice that may in particular be appropriate when study sizes, prevalences or standard errors are roughly constant. A similar approach was advocated by Shuster \emph{et~al} (2010) for general application [\cite{Shuster2010}].
  \item  \emph{Interaction weights SWADA -- weights based on the contrast estimates (previously mentioned interaction weights)}, $w_j=\frac{(\sigma_{\mathrm{A}j}^{2} + \sigma_{\mathrm{B}j}^{2})^{-1}}{\sum_{i=1}^k (\sigma_{\mathrm{A}i}^{2} + \sigma_{\mathrm{B}i}^{2})^{-1}}$: \\
  An approach somewhat similar to the within-trial framework; this prioritizes the contrast estimate (using the ``optimal'' weights for the contrasts, and possibly suboptimal ones for subgroup-specific effects). A possible variation would be (instead of using the ``common-effect weights'') to consider the estimated interaction heterogeneity ($\widehat{\tau}_\gamma$) and define $w_j=\frac{(\sigma_{\mathrm{A}j}^{2} + \sigma_{\mathrm{B}j}^{2}+\widehat{\tau}_\gamma^{2})^{-1}}{\sum_{i=1}^k (\sigma_{\mathrm{A}i}^{2} + \sigma_{\mathrm{B}i}^{2}+\widehat{\tau}_\gamma^{2})^{-1}}$. We use this heterogeneity-adjusted version in the simulation study (Table \ref{tab:sim-methods}). \\
  \item  \emph{Study size weights SWADA -- weights proportional to studies' sample-sizes}, $w_j=\frac{n_j}{\sum_{i=1}^k n_i}$: \\
  This may be a ``natural'' compromise, in particular in cases where subgroup prevalences are at least similar across studies. Sample-size weights coincide with inverse-variance weights in case variances (squared standard errors) correspond to $\sigma_j^2 = \frac{\uisd^2}{n_j}$,
  and they correspond to equal weights if all sample sizes are equal.
  Also, unlike inverse-variance weights, sample size weights are unaffected by the heterogeneity (-estimate)~[\cite{HenmiCopas2010}].
  \item  \emph{Smaller subgroup weights SWADA -- weights proportional to the smaller of both subgroups}, $w_j=\frac{\min(n_{\mathrm{A}j}, n_{\mathrm{B}j})}{\sum_{i=1}^k \min(n_{\mathrm{A}i}, n_{\mathrm{B}i})}$:\\
  This may be a ``cautious'' alternative that would also consider subgroup balance besides total study size.
  \item \emph{Minimum of three weights SWADA -- weights proportional to the minimum of three inverse-variance weights}, \(\tilde{w}_j=\min\left\{
\frac{(\sigma_{\mathrm{A}j}^{2}+\hat\tau_{\mathrm{A}}^{2})^{-1}}{\sum_{i=1}^k(\sigma_{\mathrm{A}i}^{2}+\hat\tau_{\mathrm{A}}^{2})^{-1}},
\;
\frac{(\sigma_{\mathrm{B}j}^{2}+\hat\tau_{\mathrm{B}}^{2})^{-1}}{\sum_{i=1}^k(\sigma_{\mathrm{B}i}^{2}+\hat\tau_{\mathrm{B}}^{2})^{-1}},
\;
\frac{(\sigma_{\mathrm{A}j}^{2}+\sigma_{\mathrm{B}j}^{2}+\hat\tau_{\gamma}^{2})^{-1}}{\sum_{i=1}^k(\sigma_{\mathrm{A}i}^{2}+\sigma_{\mathrm{B}i}^{2}+\hat\tau_{\gamma}^{2})^{-1}}
\right\},
\qquad
w_j=\frac{\tilde w_j}{\sum_{i=1}^k \tilde w_i}.\):\\
  For each study determine minimum across three (two subgroups, one contrast) inverse-variance weights, then re-normalize so they again sum up to 100\%. A variation similar to the previous.
  \item \emph{Minimum of total variance SWADA -- minimum total variance weights}:
  While the classical inverse-variance weights provide minimum variance for one of the three (subgroup or interaction) averages, this approach aims to minimize the overall variance of all three estimates ($\widehat{\beta}_\mathrm{A}$, $\widehat{\beta}_\mathrm{B}$ and~$\widehat{\gamma}$) simultaneously.
  The optimization basically consists of minimizing the determinant of the estimates' covariance~$P_{\widehat\beta} S P_{\widehat\beta}^{\prime}$, in which the projection matrix~\(P_{\widehat\beta}\) is constrained to equal weightings for all three averages to ensure algebraic coherence, that is the condition in Section \ref{sec:CommonWeightEstimators}. 
\end{enumerate}
Some of the above schemes are somewhat heuristic, while others are based on some kind of ``optimality'' (e.g., minimum variance for a particular estimator), and  are not going to perform well in certain situations; for example, Interaction weights SWADA yields non-zero weights only for studies that include both subgroups. In contrast, methods such as Minimum of three weights SWADA tend to favour studies with larger, more balanced samples.

In the following simulation study, we evaluate approaches introduced in this work as well as in previous literature for estimating subgroup-specific and interaction effects. We systematically compare these methods across a range of scenarios, with a primary focus on the apparent contradictions that may arise between subgroup and interaction estimates, as well as the different estimators' bias, coverage and precision. 
The goal is to assess estimation performance of each method in terms of interaction and subgroup-specific effects estimation simultaneously. To this end, we simulate and aggregate data from the Riley's IPD model [\cite{RileyEtAl2020}] and apply the methods listed in  Table \ref{tab:sim-methods}.


\section{Simulation study}\label{sec:sim}
\subsection{Aims}\label{sec:aims}

The quantities of interest are the within-trial  interaction, denoted by $\gamma$, and the expected subgroup-specific treatment effects, given by $\beta_{\mathrm{A}} = \varphi + \delta p$ and $\beta_\mathrm{B} = \varphi + \delta p + \gamma$, where $E[p_j] = p$ represents the expected subgroup prevalence (see Appendix \ref{app:TheIPDprevalenceadjustedmodelbyRileyetal}).
The data-generator model has drifting trial-specific effects with varying prevalences \(p_j\), but neither the existing approaches (e.g., the within-trial framework) nor the SWADA constraint assumes an outcome-prevalence model; they rather have a constant mean \(\varphi\) instead of \(\varphi + \delta p_j\).
When \(p_j\) varies, \(\widehat{\varphi}\) becomes a weighted average of \(\varphi+\delta p_j\). Therefore, pooled subgroup-specific effects depend on the study-level prevalence distribution and on the weighting scheme, even if the interaction is estimated within-trial (Appendix Figures~\ref{fig:Simulation_SubgroupEffectsWidth},~\ref{fig:Simulation_SubgroupEffectsBias},~\ref{fig:Simulation_SubgroupEffectsRMSE}, and \ref{fig:Simulation_SubgroupEffectsPrecision}).

Here, however, the quantities we pool are trial-level estimands (functions of each trial's subgroup composition / covariate distribution). Therefore statements about dependence on the covariate mean describe between-trial/meta-analytic behaviour and should not be interpreted as a patient-specific treatment rule. This formulation enables us to investigate how compositional imbalance and variation, captured by $\text{Var}(p_j) > 0$ with and without aggregation bias in the data-generator model. 

\subsection{Data generation and performance evaluation} \label{sec:DataGeneration}
Data are generated based on the model from Appendix \ref{app:TheIPDprevalenceadjustedmodelbyRileyetal}, that is, outcomes are simulated from normal distributions  that have ~$\uisd = 4$ (motivated by log-OR outcomes) [\cite{RoeverEtAl2021}], so that standard errors are given by $\sigma_{\mathrm{A}j}=\frac{\uisd}{\sqrt{n_{\mathrm{A}j}}}$  and $\sigma_{\mathrm{B}j}=\frac{\uisd}{\sqrt{n_{\mathrm{B}j}}}$ where $n_j = n_{\mathrm{A}j} + n_{\mathrm{B}j}$~is the study sample size. Each trial is characterized by a study-specific baseline ($\alpha_j$) and a subgroup-specific effect ($\beta_{1j}$) at the IPD level [\cite{Hua2017}], both independently drawn from a uniform distribution between $0$ and $0.5$.

To account for variability across trials, we incorporate baseline heterogeneity ($\tau$) and  interaction heterogeneity ($\tau_\gamma$), assuming identical heterogeneity for the subgroup-specific effects and half that amount for interaction terms. For each study scenario, that is, each combination of these settings, we generate \(5,000\) independent data replicates. We draw each study's size from a log-normal distribution with \(\mu=5\) and \(\sigma=1\), ensuring that on the natural scale, study sizes are positively skewed [\cite{GunhanRoeverFriede2020}]. To enforce greater similarity in study sizes within the same meta-analysis, we impose a positive correlation of~\(\rho=0.85\) between (logarithmic) sample sizes, which shrinks the spread (standard deviation) \emph{within} meta-analyses by a factor of~$0.5$.

  The remaining simulation settings are:
  \begin{enumerate}
    \item \emph{Numbers of studies ($k$)}: To avoid issues related to data sparsity (few studies), the simulations will focus on reasonably ``large'' numbers of studies in the range~$k \in \{10, 15, 20\}$. While meta-analyses from Section~\ref{sec:Examples} are used to motivate the setup of simulation scenarios, larger numbers of studies are used in order to avoid confounding with heterogeneity estimation issues or due to data sparsity. 
    \item \emph{Sample subgroup prevalences (\(p_j\))}:
        Different scenarios will be examined, where each trial's subgroup prevalence~$p_j$ is first determined and then used to assign patients within the $j$th trial. For the subgroup composition within each study, we considered a balanced scenario (1:1; \(p_j=0.50\))
and an imbalanced scenario (1:3; \(p_j=0.25\)) with identical (constant) or varying prevalence according to the following distributions. Within each trial, treatment allocation was balanced (1:1).
    \begin{itemize}
        \item Constant prevalence across trials:
            \begin{itemize}
                  \item identical / balanced (1:1, $p_j=0.50$)
                  \item identical / imbalanced (1:3, $p_j=0.25$)
            \end{itemize}
        \item  Varying prevalences across trials:
            \begin{itemize}
                  \item little variation ($p_j \sim \unifdistn(0.3, 0.7)$)
                  \item more variation ($p_j \sim \unifdistn(0.1, 0.9)$)
                  \item skewed variation ($p_j \sim \mathrm{triangular}(0.1, 0.5)$) (\emph{left-triangular} distribution, mode at~$0.1$)
       \end{itemize}
    \end{itemize}
    \item \emph{Aggregation bias (\(\delta\))}: We consider two scenarios: (i) an aggregation bias–free case, in which data are generated with $\delta = 0$ (i.e., $\gamma = \gamma_{\mathrm{Agg}} = -1$); and (ii) an aggregation bias case, in which varying prevalences influence the trial-specific treatment effect via $\gamma = -1$ and $\gamma_{\mathrm{Agg}} = 2$, yielding $\delta = 3$. 
    \item \emph{Heterogeneity values (\(\tau, \tau_\gamma\))}: We considered baseline between-study heterogeneity levels
\(\tau \in \{0,\,0.1,\,0.2,\,0.5\}\) (equivalently \(\tau^{2}\in\{0,\,0.01,\,0.04,\,0.25\}\))
and interaction heterogeneity levels
\(\tau_{\gamma} \in \{0,\,0.05,\,0.1,\,0.25\}\) (equivalently \(\tau_{\gamma}^{2}\in\{0,\,0.0025,\,0.01,\,0.0625\}\)). In addition, we set the interaction heterogeneity to be half the subgroup-specific heterogeneity,
i.e., \(\tau_{\gamma}=\tau/2\), consistent with the values above. 
 \end{enumerate}

Performance of the different approaches is evaluated using bias, interval coverage, interval width, root mean squared error (RMSE), and empirical precision across simulation replicates. These measures are complementary: bias and RMSE summarize the point estimator, while coverage and interval width summarize uncertainty quantification. Interval width is therefore interpreted separately from estimator precision, as narrower intervals do not necessarily indicate lower estimator variability when coverage differs across methods. Table~\ref{tab:sim-methods} lists all methods compared in the simulation study and the labels used in the figures.

\begin{table}[!h]
\centering
\caption{Methods compared in the simulation study. The method names are used consistently across the simulation figures. ``Separate MAs (DA and AD)'' denotes two separate random-effects meta-analyses: subgroup-specific effects are pooled using the difference in averages (DA) and interaction effects are pooled using the average difference (AD); the difference between these two quantities is the mismatch examined in Figure~\ref{fig:DeltaMismatch_Analysis}. DA, difference in averages; AD, average difference}
\label{tab:sim-methods}
\begin{tabular}{p{5cm}p{8cm}}
\hline
\textbf{Method} & \textbf{Description} \\
\hline

\multicolumn{2}{l}{\textit{Previously proposed approaches}} \\[0.5ex]

Separate MAs (DA and AD) &
Two separate random-effects meta-analyses: subgroup-specific effects are pooled separately and contrasted to obtain DA, whereas within-study interaction effects are pooled directly to obtain AD
(Section~\ref{sec:separateAnalyses}). The difference between these two quantities is the mismatch illustrated in Figure~\ref{fig:DeltaMismatch_Analysis}. \\

van Houwelingen's model &
Joint bivariate random-effects model; the interaction is obtained as the contrast of the two pooled subgroup-specific effects
(Section~\ref{sec:vhas}) \\[1ex]

Within-trial framework &
Within-trial approach for subgroup-specific effects (AD interaction)
(Section~\ref{sec:within-trial}) \\

Prevalence-adjusted model &
Prevalence adjustment to account for subgroup composition
(Section~\ref{sec:WithinBetween}) \\

\multicolumn{2}{l}{\textit{Proposed in this work}} \\[0.5ex]

Equal weights SWADA &
SWADA using equal study weights
(Section~\ref{sec:choosing}) \\

Interaction weights SWADA &
SWADA using weights tailored to the interaction estimand, including the estimated interaction heterogeneity
(Section~\ref{sec:choosing}) \\

Study size weights SWADA &
SWADA using study-size-based weights
(Section~\ref{sec:choosing}) \\

Smaller subgroup weights SWADA &
SWADA using weights based on the size of the smaller subgroup
(Section~\ref{sec:choosing}) \\

Minimum of three weights SWADA &
SWADA using the minimum of three candidate weighting schemes
(Section~\ref{sec:choosing}) \\

Minimum of total variance SWADA &
Weights chosen to minimize the estimated total variance under the SWADA constraint
(Section~\ref{sec:choosing}) \\

\hline
\end{tabular}
\end{table}

\subsection{Simulation results}\label{sec:Simulation_Studies}

\subsubsection{Mismatch between subgroup and interaction estimates}

\begin{figure}[!h]
    \centering
    \includegraphics[width = 0.9\linewidth]{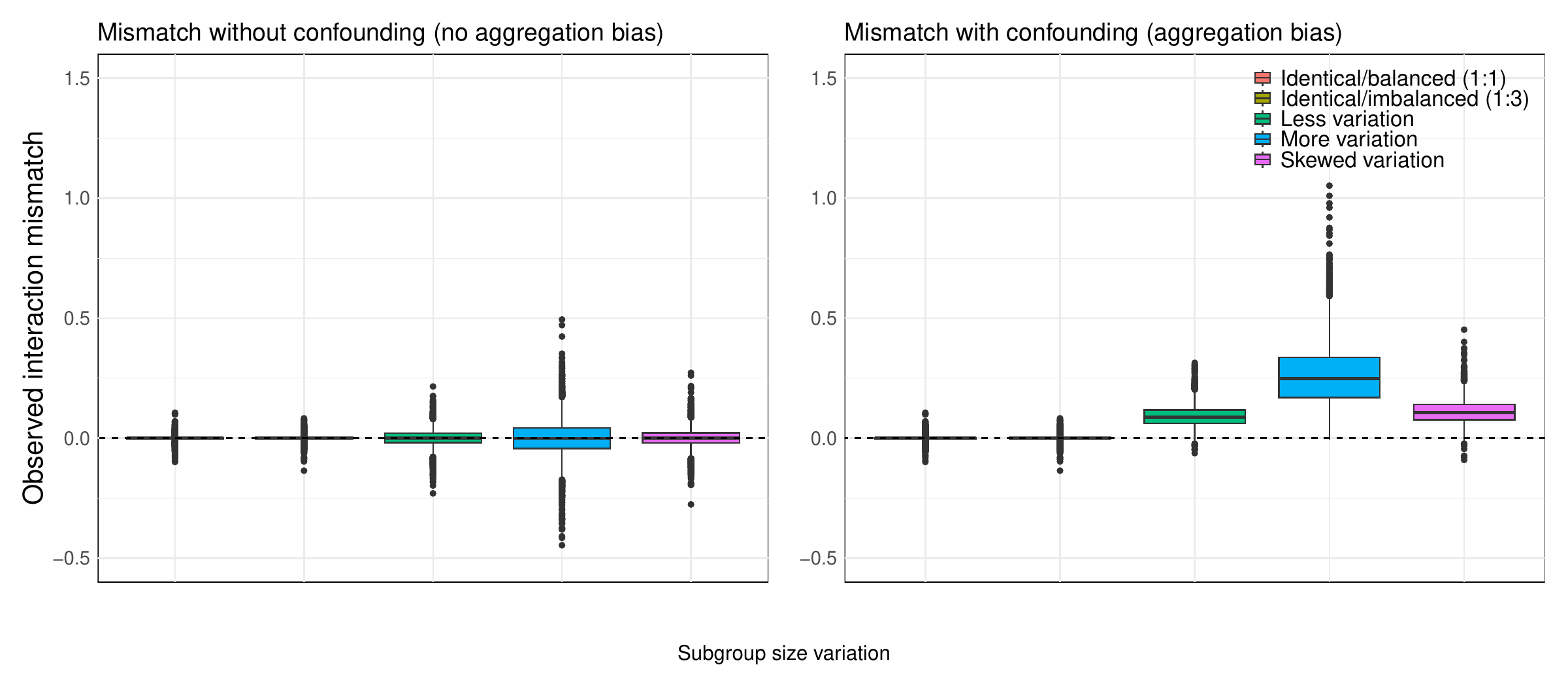}
\caption{Estimated mismatch obtained with the difference between DA and AD under a random-effects assumption. The $x$-axis represents the subgroup composition schemes defined in Section~\ref{sec:DataGeneration}, and the $y$-axis shows the magnitude of the mismatch. This scenario corresponds to an interaction heterogeneity of $\tau_\gamma = 0.1$ with $k = 20$ studies; the left-hand panel illustrates the case $\delta = 0$ (without aggregation bias), and the right-hand panel $\delta = 3$ (with aggregation bias). The right panel demonstrates that, under aggregation bias, variation in subgroup prevalence induces a systematic shift in the mismatch, while greater prevalence variation also increases its dispersion. Observed mismatches of all simulation scenarios are shown in Figure~\ref{fig:DeltaMismatch_Analysis_full}. DA, difference in averages; AD, average difference}    \label{fig:DeltaMismatch_Analysis}
\end{figure}

Figure~\ref{fig:DeltaMismatch_Analysis} illustrates the mismatch in estimates depending on whether a systematic bias was introduced during data generation ($\delta = 3$). The figure displays one representative simulation scenario with a reasonably moderate number of studies ($k = 20$). The corresponding results for the full set of simulation scenarios are presented in Appendix Figure~\ref{fig:DeltaMismatch_Analysis_full}. The overall pattern remains consistent: a noticeable location drift in the estimates emerges when the systematic association is introduced, indicating the presence of aggregation bias. Interestingly, when no prevalence variation is included, the association has no effect on the right-hand side, as matching is guaranteed. As subgroup-prevalence variation is introduced, the mismatch can become substantially larger, with the largest discrepancies in the more-variable scenarios, reflecting how trial composition amplifies the impact of the systematic association. On the other hand, the left side demonstrates that mismatch in \eqref{eqn:delta} can still arise purely from differences in subgroup composition, even in the absence of aggregation bias, while the observed location drift is specifically attributable to the latter.

\subsubsection{Coverage}
\label{sec:Coverage}
Figure~\ref{fig:Simulation_InteractionEffects} shows how often the 95\% confidence intervals contain the true interaction effect. We vary interaction heterogeneity and subgroup composition (how unevenly subjects are distributed). In the top panel, where there is no aggregation bias (no systematic outcome-prevalence link), most methods stay very close to the nominal coverage. By contrast, the bottom panel introduces a systematic prevalence–outcome trend (aggregation bias). For the van Houwelingen's model estimator, coverage drops below nominal in the scenarios with stronger prevalence variation (more/skewed variation), whereas interaction-based methods remain close to nominal. 

Figure~\ref{fig:Simulation_SubgroupEffects} summarizes the subgroup‐specific coverage. Under scenarios without aggregation bias, most methods show moderate overcoverage, indicating conservative confidence intervals, with the \emph{within‐trial framework} achieving near‐nominal coverage. When aggregation bias is introduced, coverage patterns change unevenly: several SWADA variants, particularly the \emph{equal} and \emph{minimum of three SWADA}, retain or even improve their nominal coverage, whereas others, such as the \emph{smaller subgroup SWADA}, \emph{interaction SWADA}, and \emph{study size SWADA} variants show undercoverage in some scenarios. This pattern likely reflects the inherent conservativeness of the \emph{equal} and \emph{minimum of three SWADA} weighting schemes.

In contrast, the \emph{within-trial framework}, the van Houwelingen's model, \emph{prevalence-adjusted model}, \emph{Separate MAs} and  \emph{minimum of total variance SWADA} experience pronounced declines in nominal coverage, indicating insufficient adjustment for compositional differences. Several SWADA-based estimators showed more stable coverage across bias scenarios, though often at the cost of wider intervals or reduced use of single-subgroup information. Table~\ref{tab:subgroup_coverage} summarises these subgroup-specific coverage results across all 60 simulation scenarios, showing the percentages of scenarios with under-, nominal, and overcoverage separately for settings without and with aggregation bias.

\begin{table}[!h]
\centering
\caption{Percent of the 60 simulation scenarios in which the empirical coverage of the 95\% confidence interval for the subgroup-specific effect was below 93.6\% (under), within 93.6\%--96.4\% (nominal), or above 96.4\% (over), shown separately for scenarios without and with aggregation bias. Thus, the table entries summarise the classification of the empirical coverage across the 60 data-generating mechanisms rather than the raw empirical coverage values themselves. ``Nominal'' was defined using the prespecified approximate 95\% Monte Carlo sampling interval for 1,000 replications, empirical coverage was estimated using 5,000 replications per scenario. Methods are ordered and grouped as in Table~\ref{tab:sim-methods}. DA, difference in averages; AD, average difference}
\label{tab:subgroup_coverage}

\begin{tabular}{lrrrrrr}
\toprule
 & \multicolumn{3}{c}{without aggregation bias}
 & \multicolumn{3}{c}{with aggregation bias} \\
\cmidrule(lr){2-4} \cmidrule(lr){5-7}
Method & under & nominal & over & under & nominal & over \\
\midrule

\multicolumn{7}{l}{\textit{Previously proposed approaches}} \\
Separate MAs (DA and AD)
  & 0.0 & 93.3 & 6.7 & 60.0 & 40.0 & 0.0 \\
van Houwelingen's model
  & 0.0 & 76.7 & 23.3 & 56.7 & 38.3 & 5.0 \\
Within-trial framework
  & 0.0 & 96.7 & 3.3 & 60.0 & 40.0 & 0.0 \\
Prevalence-adjusted model
  & 0.0 & 90.0 & 10.0 & 60.0 & 38.3 & 1.7 \\

\addlinespace
\multicolumn{7}{l}{\textit{Proposed in this work}} \\
Equal weights SWADA
  & 0.0 & 60.0 & 40.0 & 8.3 & 71.7 & 20.0 \\
Interaction weights SWADA
  & 0.0 & 50.0 & 50.0 & 15.0 & 66.7 & 18.3 \\
Study size weights SWADA
  & 0.0 & 38.3 & 61.7 & 16.7 & 60.0 & 23.3 \\
Smaller subgroup weights SWADA
  & 0.0 & 36.7 & 63.3 & 20.0 & 36.7 & 43.3 \\
Minimum of three weights SWADA
  & 0.0 & 55.0 & 45.0 & 1.7 & 80.0 & 18.3 \\
Minimum of total variance SWADA
  & 0.0 & 100.0 & 0.0 & 60.0 & 40.0 & 0.0 \\

\bottomrule
\end{tabular}
\end{table}

\subsubsection{Interval width and estimator precision}

\begin{figure}[!h]
    \centering
    \includegraphics[width=0.9\linewidth]{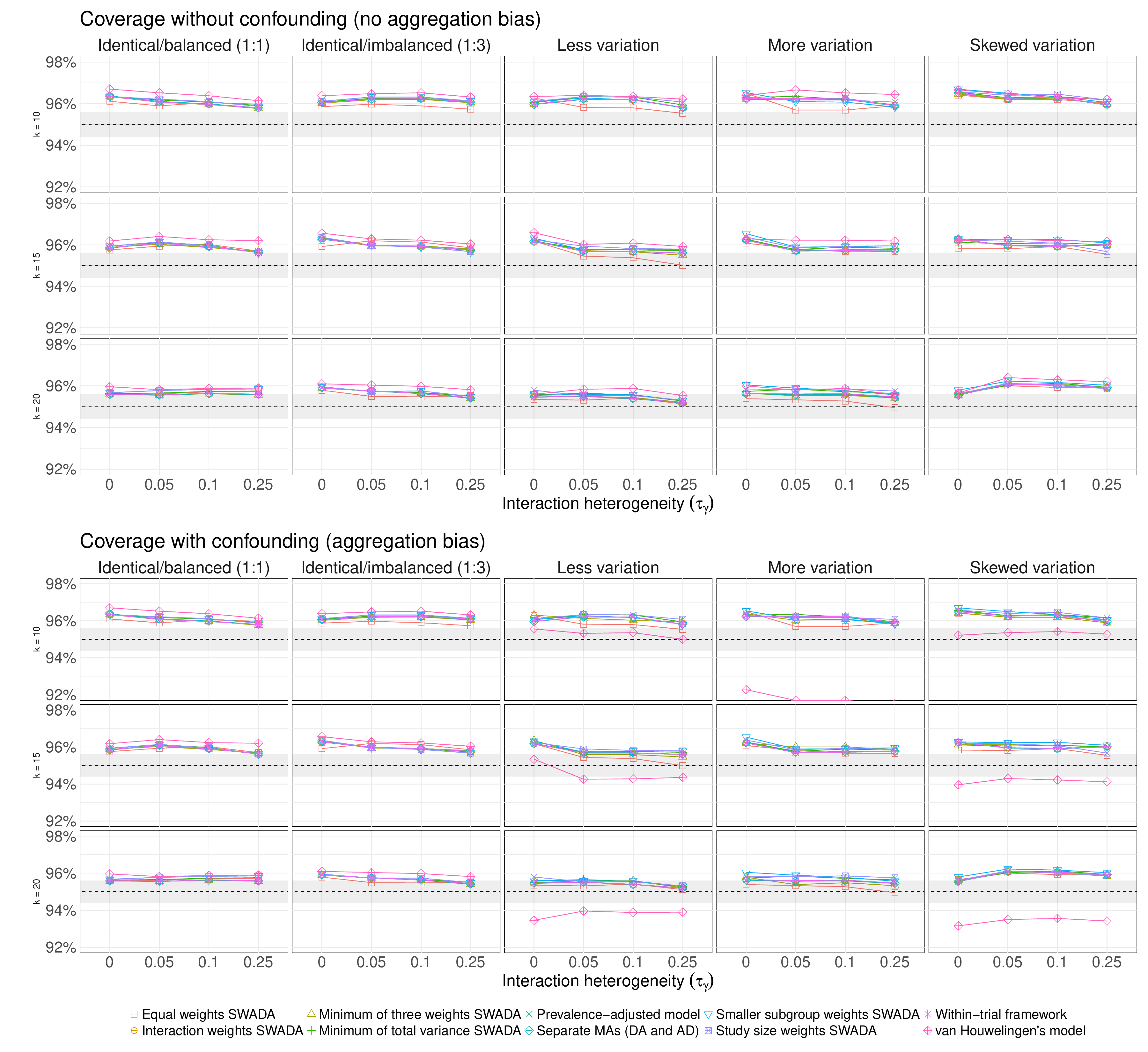}
    \caption{Coverage proportion for  treatment-by-subgroup interaction estimators in meta-analyses using different methodological approaches (for a nominal level of~$95\%$). The top panels show performance on data generated with no aggregation bias and the bottom panels under an aggregation bias assumption. The dashed horizontal line indicates the nominal 95\% coverage, and shaded areas represent Monte Carlo confidence intervals. Each column corresponds to a different subgroup-prevalence (subgroup-composition) scenario (Section \ref{sec:DataGeneration}): identical/balanced (1:1); \(p_j=0.50\), identical/imbalanced (1:3); \(p_j=0.25\), little variation \(p_j\sim \mathrm{Uniform}(0.3,0.7)\), more variation \(p_j\sim \mathrm{Uniform}(0.1,0.9)\), and skewed variation \(p_j\sim \mathrm{triangular}(0.1,0.5)\), mode 0.1). Because this figure concerns the interaction estimand, ``Separate MAs (DA, difference in averages; AD, average difference)'' appears in its AD form (within-study interactions pooled directly); the subgroup-effect counterpart (DA) is shown in Figure~\ref{fig:Simulation_SubgroupEffects}}\label{fig:Simulation_InteractionEffects}
   \centering     
\end{figure}

\begin{figure}[!h]
    \centering
    \includegraphics[width=0.9\linewidth]{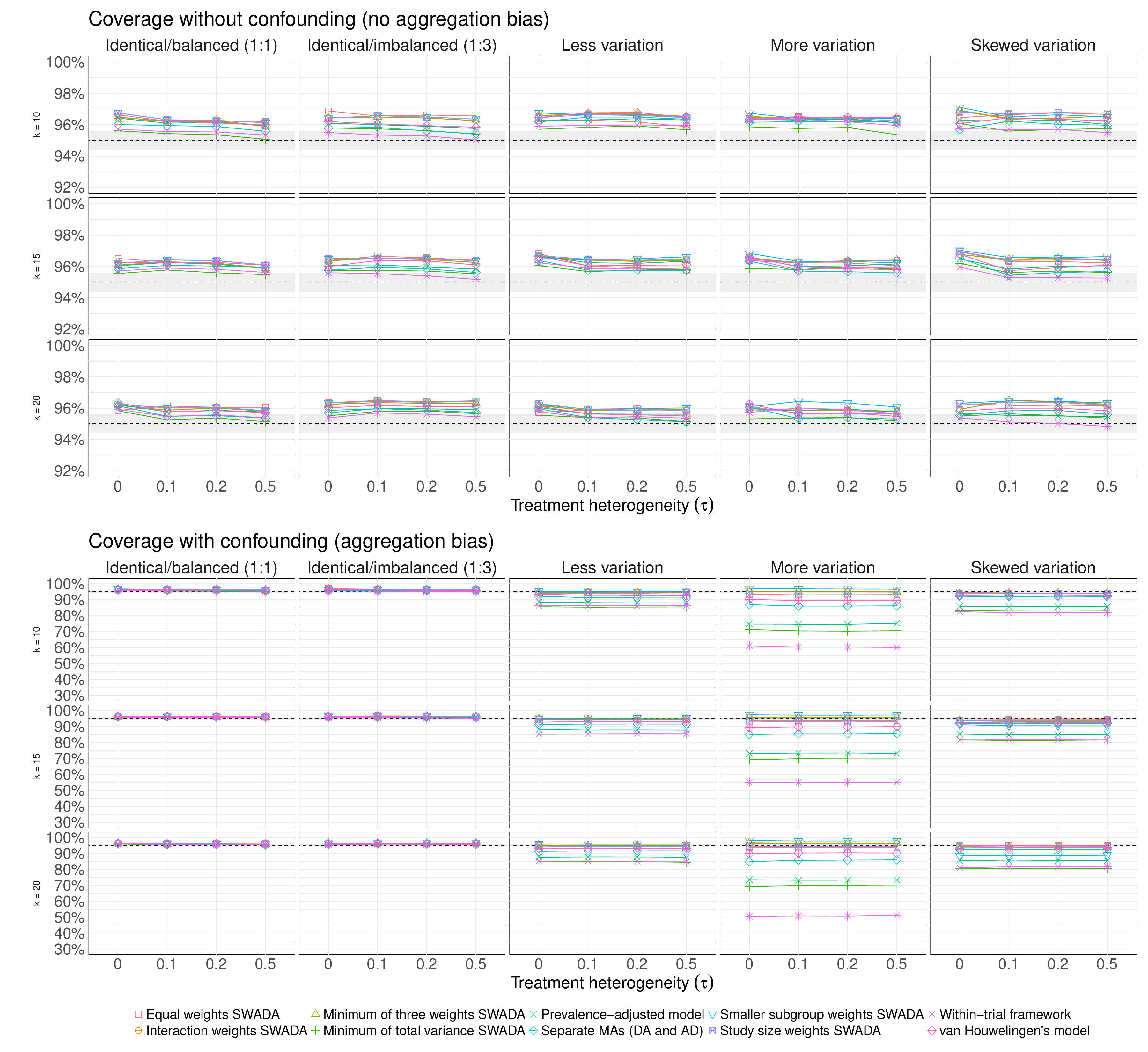}
\caption{Coverage proportion for the expected reference subgroup's treatment effect under various methodological approaches (for a nominal level of~$95\%$). Here, the reference-subgroup-specific effect refers to $\beta_A$ in Section~\ref{sec:aims} (treatment effect at the average covariate value \(p_j = p\)), not the covariate-averaged subgroup-specific effects. The top panels show performance on data generated with no aggregation bias and the bottom panels under an aggregation bias assumption. The dashed horizontal line indicates the nominal 95\% coverage, and shaded areas represent Monte Carlo confidence intervals. Each column corresponds to a different subgroup-prevalence (subgroup-composition) scenario (Section~\ref{sec:DataGeneration}): identical/balanced (1:1), \(p_j=0.50\); identical/imbalanced (1:3), \(p_j=0.25\); little variation, \(p_j\sim \mathrm{Uniform}(0.3,0.7)\); more variation, \(p_j\sim \mathrm{Uniform}(0.1,0.9)\); and skewed variation, \(p_j\sim \mathrm{triangular}(0.1,0.5)\), mode 0.1.}
\label{fig:Simulation_SubgroupEffects}
\end{figure}

Figure~\ref{fig:Simulation_InteractionEffectsWidth} presents the width ratios of interaction interval estimates relative to the standard or reference AD, expressed as $100 \times \frac{SE_{\mathrm{method}}}{SE_{\mathrm{AD}}}$. Zooming into the scenario depicted in Figure~\ref{fig:DeltaMismatch_Analysis} ($k = 20$, ``More variation''), the \emph{equal weights SWADA} yields the widest intervals, near 14\% larger on average than the reference, while the \emph{study size SWADA} and \emph{smaller subgroup SWADA} also tend to be wide, yielding intervals that are 2–5\% larger than AD. By design, the \emph{within‐trial framework}, the \emph{prevalence‐adjusted}, and the \emph{interaction weights SWADA} methods all have the reference width (ratio = 100\%). Although the non‐linear estimators can sometimes produce shorter interaction intervals, van Houwelingen’s model can exhibit some bias (Appendix Figure~\ref{fig:Simulation_InteractionEffectsBias}) together with shorter intervals (Appendix Figure~\ref{fig:Simulation_InteractionEffectsWidth}) in these scenarios, both of which may contribute to its lower coverage as shown in Figure~\ref{fig:Simulation_InteractionEffects}.

Figure~\ref{fig:Simulation_SubgroupEffectsWidth} shows the analogous ratios for the subgroup-specific effect estimator versus the standard DA. Under the same scenario shown in Figures~\ref{fig:DeltaMismatch_Analysis} and~\ref{fig:Simulation_InteractionEffectsWidth}, most methods  yield interval widths within about \(\pm20\%\) of the DA reference width with no aggregation bias, while existing methods, including \emph{within-trial framework} and \emph{prevalence-adjusted model} can produce substantially narrower intervals, in the most extreme scenarios approaching half the DA width. These results hold across all design settings (albeit on different scales) and show only modest growth in these effects as heterogeneity increases (see Figure~\ref{fig:Simulation_SubgroupEffectsWidth}). Some SWADA-based intervals are accompanied by conservative coverage, typically at or just above the nominal 95\% level. Figures~\ref{fig:DeltaMismatch_Analysis_full}--\ref{fig:Simulation_SubgroupEffectsPrecision} show comprehensive results across the whole range of studied scenarios. 

Figure~\ref{fig:Simulation_InteractionEffectsWidth} shows interval widths for estimating the interaction effect \(\gamma\), and Figure~\ref{fig:Simulation_SubgroupEffectsWidth} shows the corresponding interval widths for the subgroup-specific effect. Because interval width reflects the reported uncertainty rather than estimator variability itself, these results should be interpreted together with the bias and RMSE results reported in Figures~\ref{fig:Simulation_InteractionEffectsBias}–\ref{fig:Simulation_SubgroupEffectsRMSE}. For the interaction estimator, precision is generally higher when subgroup prevalences are identical or only mildly variable, and decreases as between-study subgroup composition becomes more variable, especially under aggregation bias. Equal-weights SWADA tends to have the lowest precision across scenarios, whereas van Houwelingen's model and the within-trial framework often show relatively high precision in the absence of aggregation bias, although this advantage does not persist in all aggregation-bias scenarios. For the reference subgroup-specific effect, precision increases with the number of studies and, in the absence of aggregation bias, is generally highest in the identical/imbalanced and skewed-prevalence scenarios; under aggregation bias, stronger prevalence variation substantially reduces precision. These patterns indicate that weighting choices affect not only bias and coverage, but also estimator efficiency. As an additional summary of point-estimator performance, we also report RMSE for the treatment-by-subgroup interaction and reference subgroup-specific estimators in Figures~ \ref{fig:Simulation_InteractionEffectsRMSE} and \ref{fig:Simulation_SubgroupEffectsRMSE}. The RMSE results were broadly consistent with the bias and empirical-precision findings: methods with lower bias did not always achieve the smallest RMSE when estimator variance increased. This further illustrates the trade-off underlying SWADA, where enforcing algebraic agreement may improve coherence but can reduce efficiency under some weighting schemes. 

\section{Example applications revisited}\label{sec:example}

Simulation results suggested that some methods may offer better performance under specific conditions. Unlike the revisited meta-analysis, the current simulation study did not explicitly incorporate single-subgroup trials such as RECOVERY, since the routine inclusion of such large, single-subgroup trials is context-dependent (Section \ref{sec:Dealingwithsinglesubgrouptrials}). Consequently, the performance of each estimator under extreme single‐subgroup is discussed for this particular example. 

\begin{figure}[!h]
\centering
\includegraphics[width= 0.9\linewidth]{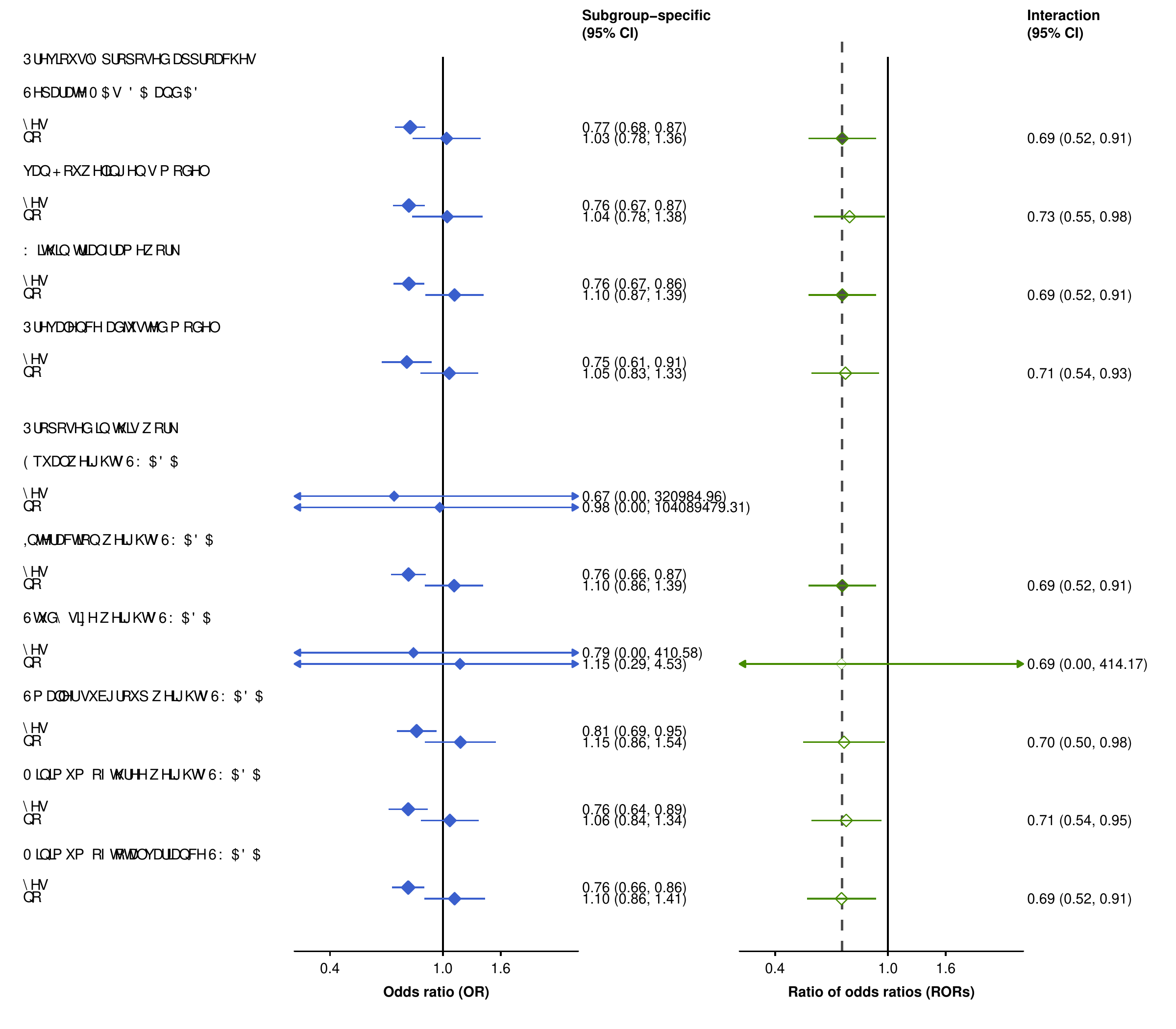}
\caption{Forest plot comparing different methods for estimating the effect of IL-6 antagonists treatment on mortality in COVID-19 patient subgroups (by corticosteroid usage; \emph{yes} vs. \emph{no}) along with the  interaction (the difference between subgroups)~(see also Figure~\ref{fig:IL6_COVID19_Analysis}).
The left panel shows combined subgroup-specific treatment effects (odds ratios, ORs) for patients not requiring corticosteroid administration (no) and those requiring corticosteroid (yes) for various methods.
The right panel illustrates the associated  interaction effects (ratios of odds ratios, RORs) comparing the ``steroid-yes'' and ``steroid-no'' groups. The  inverse-variance weights average difference (AD) estimate is also indicated by a vertical dashed line, and estimators that will \emph{by construction} yield the same interaction estimate are shown with a filled diamond. DA, difference in averages; AD, average difference}
\label{fig:Final_Results_IL6}
\end{figure}

\begin{figure}[!h]
\centering
\includegraphics[width = 0.9\linewidth]{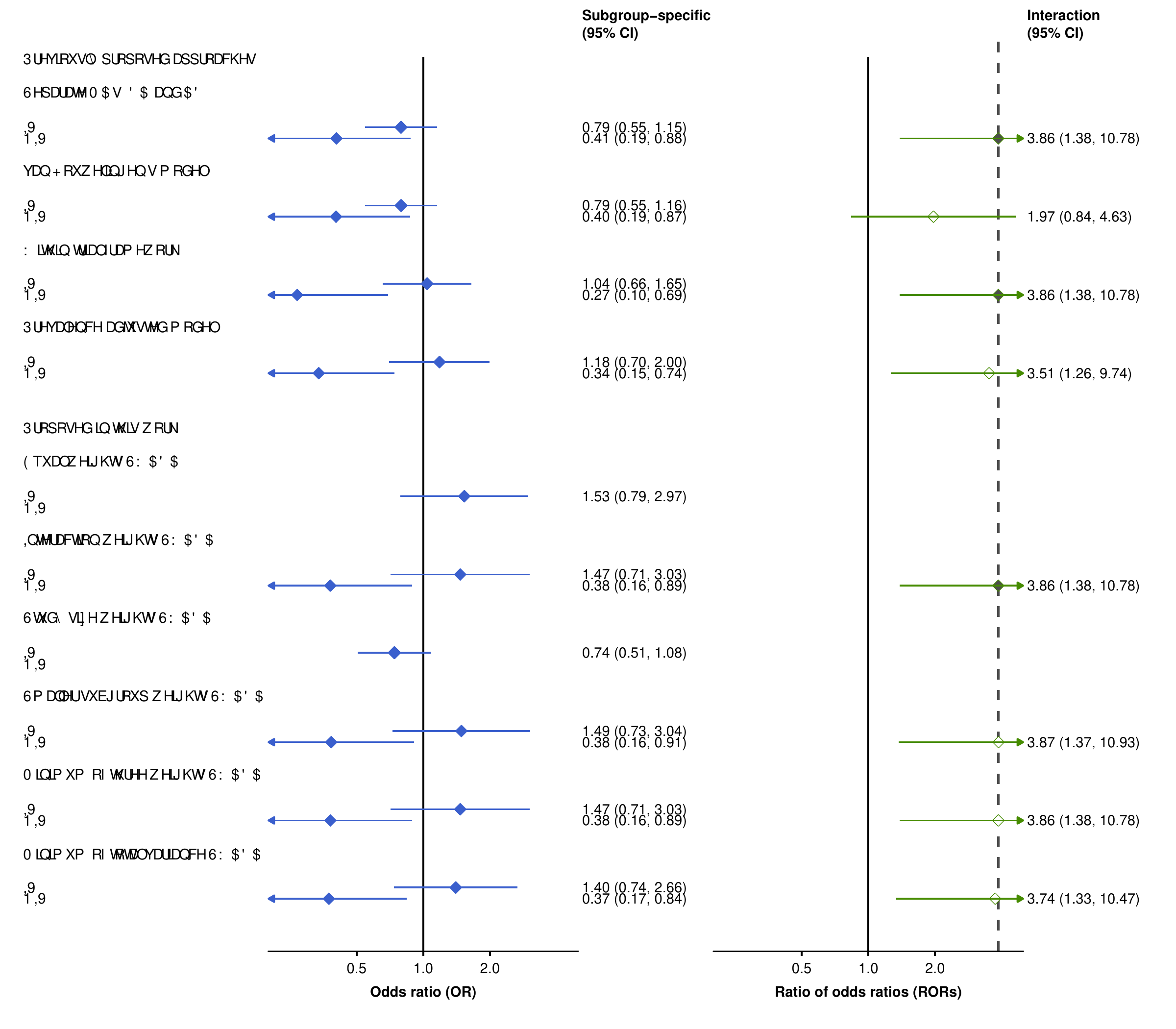}
\caption{Forest plot comparing different methods for estimating the effect of corticosteroid treatment on mortality in patient subgroups (by invasive ventilation status; \emph{IV} vs. \emph{NIV}) along with the  interaction (the difference between subgroups) in COVID-19 patients (see also Figure~\ref{fig:Corticosteroids_MetaAnalysis}).
The left panel shows combined subgroup-specific treatment effects (odds ratios, ORs) for patients requiring invasive ventilation (IV) and those not requiring invasive ventilation (NIV) for various methods.
The right panel illustrates the corresponding  interaction effects (ratios of odds ratios, RORs) comparing the IV and NIV groups. The  inverse-variance weights  average difference (AD) estimate is also indicated by a vertical dashed line, and estimators that will \emph{by construction} yield the same interaction estimate are shown with a filled diamond. DA, difference in averages; AD, average difference} \label{fig:Final_Results_Corticosteroids}
\end{figure}

In the original Corticosteroids meta-analysis, despite possible heterogeneity arising from both clinical diversity and methodological variations [\cite{CochraneHandbook}], the original analysis took an optimistic stance and assumed between-trial homogeneity of subgroup-specific and interaction effects. 
As noted in Section \ref{sec:Examples}, heterogeneity is sometimes found to be zero for separate analyses. Similar behaviour has been discussed extensively in the literature: Röver \emph{et~al} [\cite{RoeverKnappFriede2015}] highlighted the sensitivity of frequentist meta-analyses with few studies and the role of the Knapp–Hartung adjustment; Friede \emph{et~al} [\cite{FriedeRoeverWandelNeuenschwander2017a, FriedeRoeverWandelNeuenschwander2017b}] demonstrated how Bayesian modelling can provide more robust inference in such settings; and Röver \emph{et~al} [\cite{RoeverRindskopfFriede2024}] illustrated diagnostic aspects, such as the use of trace plots to assess the random-effects model adequacy. In this work we re-analysed the data under the preferred random-effects assumption, previously shown in Figures \ref{fig:IL6_COVID19_Analysis} and \ref{fig:Corticosteroids_MetaAnalysis}. Figure~\ref{fig:Final_Results_IL6} shows the results of the methods described in Table \ref{tab:sim-methods} for the IL-6 antagonist meta-analysis data, while Figure~\ref{fig:Final_Results_Corticosteroids} presents the corresponding pooled estimates across methods.

Single-subgroup contributions such as RECOVERY in the second example (Figure \ref{fig:Corticosteroids_MetaAnalysis}) tend to exaggerate  the contradiction. The extremely wide confidence intervals for equal-weights and study-size-weights SWADA in Figure \ref{fig:Final_Results_IL6} are driven by the single-subgroup trials. For these trials, the unavailable subgroup is represented by an effectively uninformative estimate with a very large variance; because equal and study-size weighting do not down-weight such studies according to this lack of precision, the large variance is propagated into the pooled estimates. For equal-weights SWADA, the resulting uncertainty in the interaction estimate is so extreme that no informative finite confidence interval can be displayed. Appendix Figures \ref{fig:Final_Results_IL6_excluded} and \ref{fig:Final_Results_Corticosteroids_excluded} show results after the single-subgroup trials exclusion. Notably this exclusion dramatically changes the point estimates of existing methods, whereas some SWADA weightings change little because single-subgroup studies receive negligible or zero weight in these estimators, resulting in a ``virtual exclusion'' mechanism. Other SWADA variants may show sensitivity, in particular, study size SWADA.

Several SWADA variants maintained coverage better in the simulation settings; for some variants, this came with wider intervals, and some weighting choices also effectively excluded single-subgroup trials. Whether a SWADA estimate may be suitable for a primary analysis hence very much depends on the context, in particular, on the expected degree of subgroup imbalance, and the necessity for matching subgroup effect and interaction estimates. Regarding the Corticosteroid meta-analysis example (Figure \ref{fig:Final_Results_Corticosteroids}), the following considerations may play a role:
\begin{enumerate}
    \item Which methods do/do not reproduce exactly the AD estimates (point and interval estimates), i.e. within-trial interaction ROR \(\approx 3.9\) 
    \begin{itemize}
        \item Do:  Three approaches: the Within‐trial framework, interaction weights SWADA and Separate MAs (that is again AD). All yield an identical ROR of \(3.86~(\text{95\% CI}~ 1.38 - 10.78)\).
            \item Do not: The van Houwelingen's model gives a smaller ROR of \(1.97~(0.84 - 4.63)\). Prevalence-adjusted model sits in between at \(3.51~(1.26 - 9.74)\), and other SWADAs. Those showed similar uncertainty but did not exactly recover the AD-ROR of ~3.8  with slightly different estimates (\(3.74\) and \(3.87\)). Study size and Equal weights SWADAs have produced unstable estimates in the presence of single-subgroup trials.
    \end{itemize}
    \item Which ``matching'' methods drive implausible NIV subgroup ORs
    \begin{itemize}
        \item Among the three options that lock onto ROR \(\approx 3.86\), the within-trial framework does so by shrinking the NIV OR to \(0.27\) (\(\text{95\% CI} \) ~\(0.10 - 0.69)\), slightly more protective than the smallest individual NIV point estimate shown in 0.28–2.00. Interaction weights SWADA reproduces the same interaction estimate while yielding a more moderate NIV OR; however, it does so by effectively down-weighting large single-subgroup trials such as RECOVERY, and may therefore be less suitable when the subgroup-specific effect in that subgroup is itself of interest.
        \item Additionally, several other SWADA-based estimates that do not exactly recover the AD-ROR produce more moderate NIV ORs around \(0.40\).
    \end{itemize}
     \item Precision trade-offs:  
     Joint subgroup analyses that focus on the ``AD-ROR'' produce confidence intervals spanning approximately $7.8$–$8.7\times$, compared with $5.5\times$ under DA, reflecting the corresponding loss of precision (see also Figure \ref{fig:Simulation_SubgroupEffectsWidth}).
\end{enumerate}

\section{Discussion}\label{sec:dis}

Understanding and accurately estimating treatment effects across different subgroups in meta-analyses is crucial for informing policy and clinical decision-making. The challenge lies in managing the inherent heterogeneity and interaction effects that arise when combining studies with varying subgroup prevalences and treatment interactions. This study aimed to explore the performance of different meta-analytic methods under such conditions, offering insights into the trade-offs between precision and coverage in estimating  interactions. Coverage was generally close to nominal in the absence of aggregation bias and remained relatively stable for most interaction estimators. Larger departures occurred for subgroup-specific estimators, particularly when substantial variation in subgroup prevalence was coupled with outcome-prevalence associations, leading to undercoverage. This issue is exacerbated in scenarios involving a limited number of studies, a challenge previously documented by Friede \emph{et~al} [\cite{FriedeRoeverWandelNeuenschwander2017a}]. The limited study numbers in these cases often result in inadequate power and increased susceptibility to errors, particularly when traditional meta-analytic techniques are applied. Using subgroup information can be seen as a step towards conducting an \emph{individual participant data meta-analysis} (IPD-MA) that widens the scope of a ``simple'' meta-analysis and naturally allows for the examination of relative efficacy between subgroups [\cite{RileyEtAl2020}]. Such interactions play a pivotal role in policy or treatment decisions, determining whom to treat and how, and quantifying potential variations in benefit/risk balances for different patient groups in personalized medicine and health technology assessment [\cite{IQWiGMethodsV5, Phillippo2018}].\\

These differences are not merely technical: they carry substantial consequences for interpretation of the pooled results. In evidence synthesis guiding clinical recommendations, underestimating subgroup differences due to aggregation bias could mask treatment effects for high-risk populations, while overestimating them could wrongly influence subgroup prioritization in health policy. Coverage and interval width differed across methods: some approaches produced substantially narrower intervals but lower coverage under aggregation bias, whereas some SWADA weighting schemes achieved more stable coverage at the cost of wider intervals. This trade-off highlights the complexity of balancing precision and reliability in meta-analytic estimates when subgroup-specific effects are of interest. The study also underscores the difficulties in consistently estimating treatment effects across subgroups using standard weights or within-trial matching strategies, especially with aggregation bias. An interesting avenue for improving meta-analytic methods based on standard weights is suggested by the work of Henmi and Copas [\cite{HenmiCopas2010}]. The use of pre-specified weights, as they propose, could mitigate some of the observed issues, such as undercoverage and subgroup imbalance, thereby enhancing the reliability of meta-analytic results. By addressing these limitations, we can potentially develop more robust techniques for analysing treatment effects in the presence of significant heterogeneity and interaction effects. In the aggregation-bias scenarios considered here, the within-trial framework and standard-weight approaches showed reduced coverage for the subgroup-specific estimand. This limitation points to a broader issue in the robustness of traditional meta-analytic techniques when applied to complex, heterogeneous data.

Upon investigating the effects of different weighting schemes employed in different effect and interaction estimators, here we have proposed the use of a SWADA approach, that is, using the same weighting for both subgroup-specific effect as well as interaction estimates. Interaction weights retain the conventional within-trial interaction estimate, but are not always appropriate because they assign zero weight to single-subgroup trials and can discard substantial subgroup evidence (e.g. RECOVERY). They may also perform inefficiently, or remain undefined, when single-subgroup trials are dominant, as in the corticosteroid example application from Section~\ref{sec:ReactExampleIntro}.

Across the methods and examples, we often end up trading precision for consistency of estimates, or conversely, accepting discrepancies in order to obtain ``optimal'' intervals. The study also highlights the need for methodological improvements, such as enhancing estimation at the subgroup level and incorporating Bayesian methods, to better capture between-trial heterogeneity and improve the reliability of meta-analytic conclusions.


\section*{Acknowledgment}
We thank David Rindskopf (The Graduate Center, City University of New York) for his careful review of the manuscript and for constructive comments that improved its clarity and presentation.



\clearpage
\begin{appendix}
\section{Appendix}

\subsection{Excluding single-subgroup trials} \label{app:Excluded}

\begin{figure}[!h]
\centering
\includegraphics[width = 0.9\linewidth]{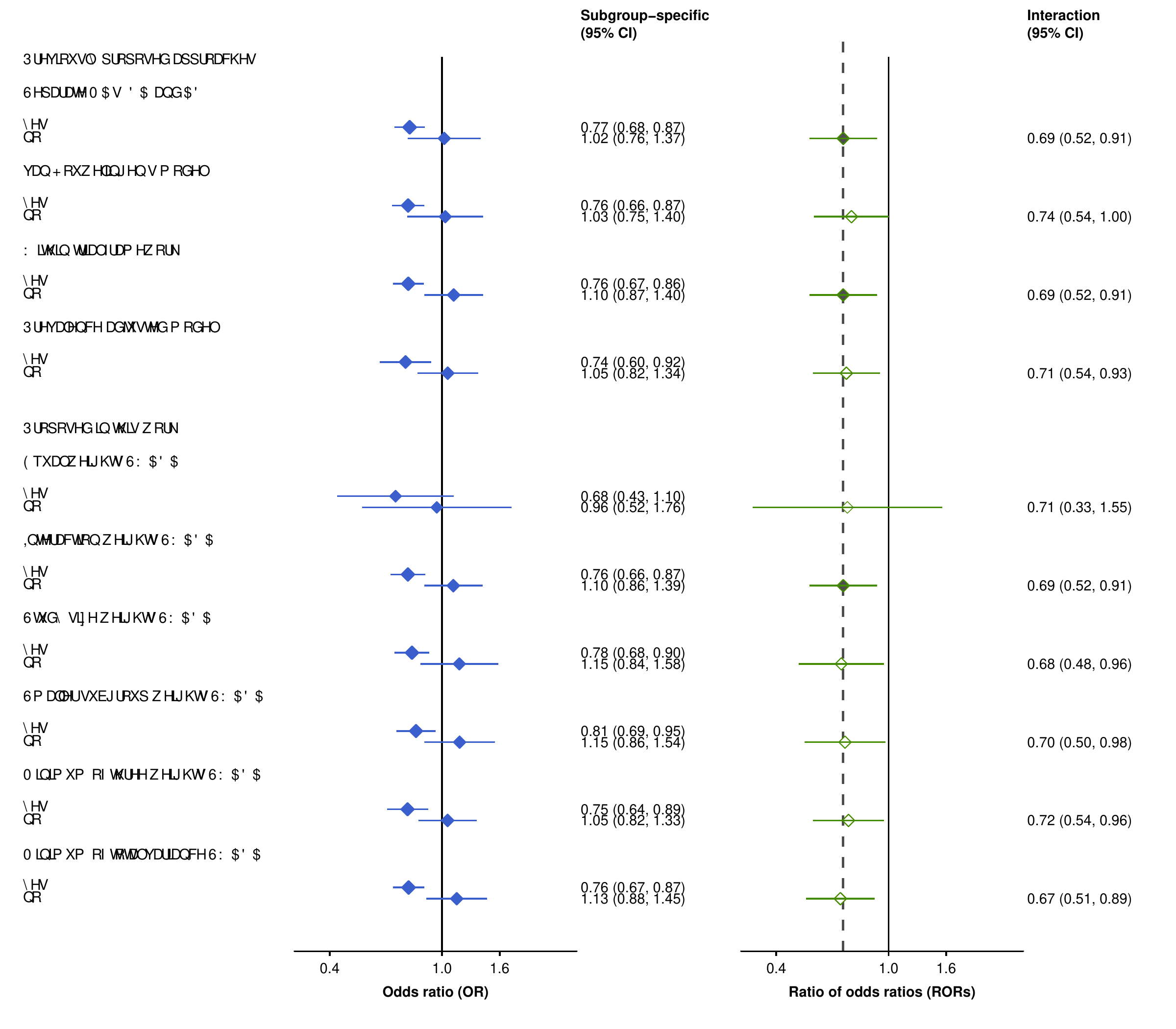}
\caption{Forest plot comparing different methods for estimating the effect of IL-6 antagonists treatment on mortality in patient subgroups (by corticosteroid usage; yes vs. no) along with the  interaction (the difference between subgroups) in COVID-19 patients (see Figures~\ref{fig:IL6_COVID19_Analysis} and \ref{fig:Final_Results_IL6}). Single-subgroup trials (reporting data for only one steroid subgroup) were excluded from this analysis. The left panel shows combined subgroup-specific treatment effects (odds ratios, ORs) for patients not requiring corticosteroid administration (no) and those requiring corticosteroid (yes) for various methods. The right panel illustrates the treatment-by-steroid interaction effects (ratios of odds ratios, RORs) comparing the ``steroid-yes'' and ``steroid-no'' groups. The inverse-variance weights  average difference (AD) is also indicated by a vertical dashed line, and estimators that will \emph{by construction} yield the same interaction estimate are shown with a filled diamond. DA, difference in averages; AD, average difference} \label{fig:Final_Results_IL6_excluded}
\end{figure}

\begin{figure}[!h]
\centering
\includegraphics[width = 0.9\linewidth]{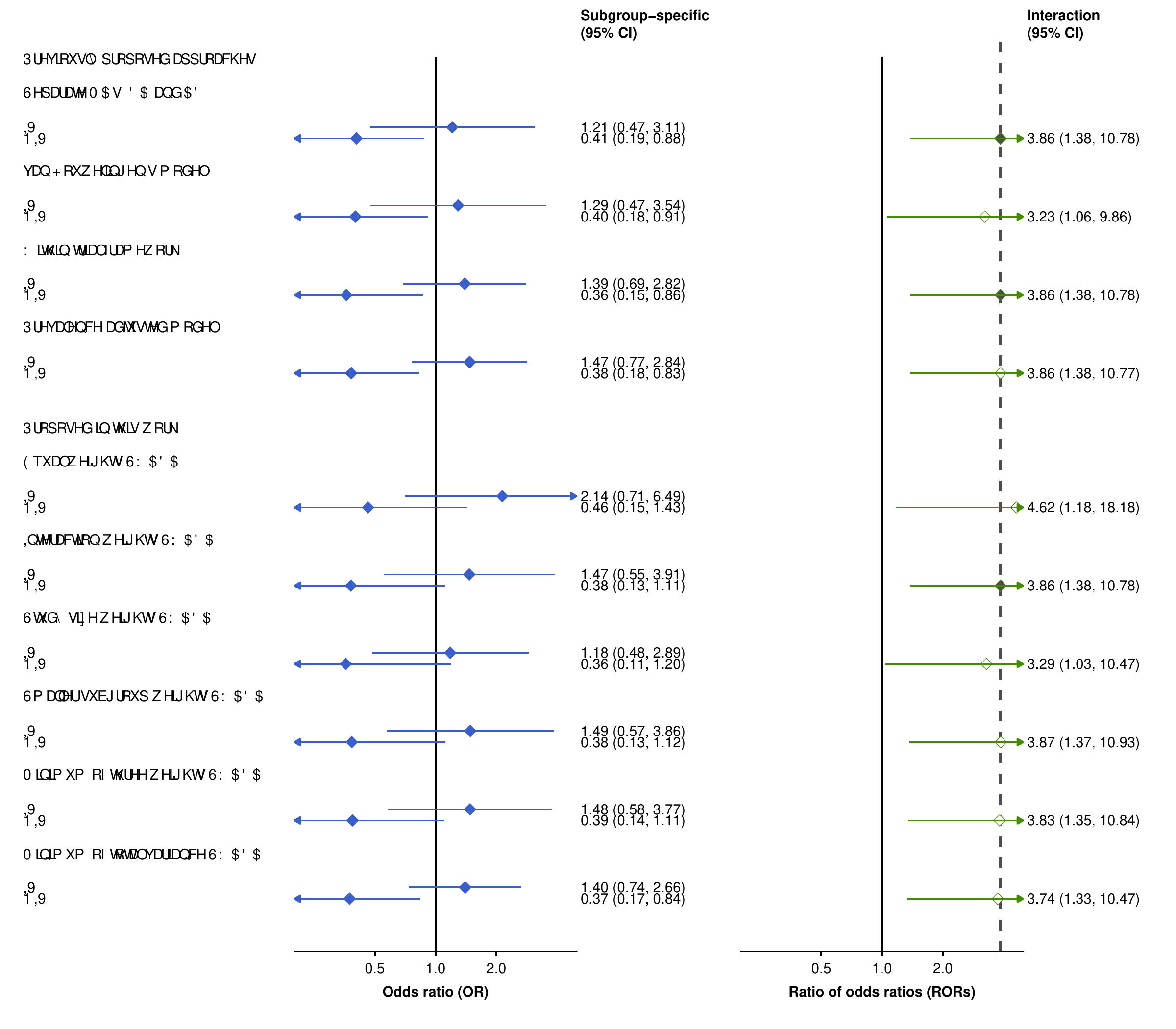}
\caption{Forest plot comparing different methods for estimating the effect of corticosteroid treatment on mortality in patient subgroups (by invasive ventilation status; IV vs. NIV) along with the treatment-by-ventilation interaction (the difference between subgroups) in COVID-19 patients (see Figures~\ref{fig:Corticosteroids_MetaAnalysis} and ~\ref{fig:Final_Results_Corticosteroids}). Single-subgroup trials (reporting data for only one ventilation subgroup) were excluded from this analysis. The left panel shows combined subgroup-specific treatment effects (odds ratios, ORs) for patients requiring invasive ventilation (IV) and those not requiring invasive ventilation (NIV) for various methods. The right panel illustrates the  treatment-by-subgroup interaction effects (ratios of odds ratios, RORs) comparing the IV and NIV groups. The inverse-variance weights average difference (AD) estimate is also indicated by a vertical dashed line, and estimators that will \emph{by construction} yield the same interaction estimate are shown with a filled diamond. DA, difference in averages; AD, average difference}
\label{fig:Final_Results_Corticosteroids_excluded}
\end{figure}

Figure \ref{fig:Final_Results_Corticosteroids_excluded}
shows results for the corticosteroids-by-ventilation meta-analysis on COVID-19 patients when single-subgroup trials are not included. Figure \ref{fig:Final_Results_IL6_excluded}
shows results for the IL-6-antagonists-by-corticosteroid-use meta-analysis on COVID-19 patients when single-subgroup trials are not included. As expected, removing single-subgroup trials primarily affects methods that pool subgroup estimates directly or rely on conventional random-effects weighting. In contrast, the interaction weights, smaller-subgroup-weighted, and minimum-of-three-weighted SWADA estimators produce virtually identical results with or without such trials. Caution remains for (total) study size weights, as RECOVERY has strong influence in the treatment-by-ventilation analysis.

\subsection{SWADA consistency under confounding}
\label{app:SWADAunbiasedness}
\subsubsection{Treatment-by-subgroup interaction effect unbiasedness}

Using the prevalence-adjusted model formulation from Section~\ref{sec:WithinBetween}, the interaction bias is expressed as
\begin{eqnarray}
\mathrm{Bias}(\widehat{\gamma})&=& \expect\!\left[\widehat{\gamma}\right] - \gamma\\
&=& 
\expect\!\left[
  \sum_{j=1}^k w_{y_{\mathrm{B}_j}} y_{B_j} 
  - \sum_{j=1}^k w_{y_{\mathrm{A}_j}} y_{A_j}
\right]
- \gamma
 \\[4pt]
&=&
\sum_{j=1}^k 
  w_{y_{\mathrm{B}_j}}\, \expect[y_{B_j}]
- \sum_{j=1}^k 
  w_{y_{\mathrm{A}_j}}\, \expect[y_{A_j}]
- \gamma
 \\[4pt]
&=&
\sum_{j=1}^k 
  w_{y_{\mathrm{B}_j}}\!\left(\varphi + \delta p_j + \gamma\right)
- \sum_{j=1}^k 
  w_{y_{\mathrm{A}_j}}\!\left(\varphi + \delta p_j\right)
- \gamma
 \\[4pt]
&=&
\delta \sum_{j=1}^k p_j \bigl(w_{y_{\mathrm{B}_j}} - w_{y_{\mathrm{A}_j}}\bigr) = 0, \quad \mathrm{since} \quad w_{y_{\mathrm{B}_j}} = w_{y_{\mathrm{A}_j}}.
\label{eqn:aggbias}
\end{eqnarray}
where \(w_{y_{\mathrm{A}j}}\) and \(w_{y_{\mathrm{B}j}}\) denote the normalised-weights assigned to study \(j\) for subgroups~A and~B, respectively; \(\varphi\) is a baseline effect; \(\delta\) quantifies the linear prevalence effect associated with the subgroup prevalence~\(p_j\) in subgroup~B; and \(\gamma_\mathrm{W}\)~is the within-trial interaction effect in \eqref{eqn:withintrialframework}.

\subsubsection{Pooled subgroup-specific effects under SWADA}
Regarding the pooled subgroup-specific effects, let  
\(\widehat{\theta}\) denote the estimator obtained by pooling the subgroup-specific effects.
\begin{eqnarray}
\mathrm{Bias}(\widehat{\theta}) &=&  \expect[\widehat\theta] - (\varphi + \delta \bar p)\\
&=& 
\frac{
  \sum_{j=1}^k 
  \big(w_{y_{\mathrm{A}_j}}\expect[y_{A_j}] + w_{y_{\mathrm{B}_j}}\expect[y_{B_j}]\big)
}{
  \sum_{j=1}^k (w_{y_{\mathrm{A}_j}} + w_{y_{\mathrm{B}_j}})
}
- (\varphi + \delta \bar p)
 \\[4pt]
&=& 
\frac{
  \sum_{j=1}^k (w_{y_{\mathrm{A}_j}} + w_{y_{\mathrm{B}_j}})(\varphi + \delta p_j)
  + \gamma\sum_{j=1}^k w_{y_{\mathrm{B}_j}}
}{
  \sum_{j=1}^k (w_{y_{\mathrm{A}_j}} + w_{y_{\mathrm{B}_j}})
}
- (\varphi + \delta \bar p)
 \\[6pt]
&=& 
\delta\,(\bar p_w - \bar p)
+\,\frac{W_B}{W}\gamma\\
&=& 
\delta\,(\bar p_w - \bar p)
+ 0.5\gamma, \quad \mathrm{since} \quad w_{y_{\mathrm{B}_j}} = w_{y_{\mathrm{A}_j}}.
\label{eqn:pooled_bias}
\end{eqnarray}
where \(W_B\) denotes the total subgroup B
weight and \(W\) the total weight, with \(
W_B = \sum_{j=1}^k w_{y_{\mathrm B_j}} = \sum_{j=1}^k w_j,
\mathrm{~and~} 
W = \sum_{j=1}^k \bigl(w_{y_{\mathrm A_j}} + w_{y_{\mathrm B_j}}\bigr)
   = 2\sum_{j=1}^k w_j,
\)
where \(\bar{p}_w = \sum_{j=1}^k w_j p_j\) denotes the \textit{weighted average prevalence}. 
In this way, under the present reference-subgroup parameterization, a SWADA-based overall treatment effect estimator has an offset of \(0.5\gamma\) relative to the reference-subgroup effect, provided that \(
\bar{p}_w - \bar{p} \approx 0
\) is negligible.

\subsection{Mismatch between inverse-variance weights when done separately}
\begin{figure}[!h]
    \centering    
    \includegraphics[width = 0.9\linewidth]{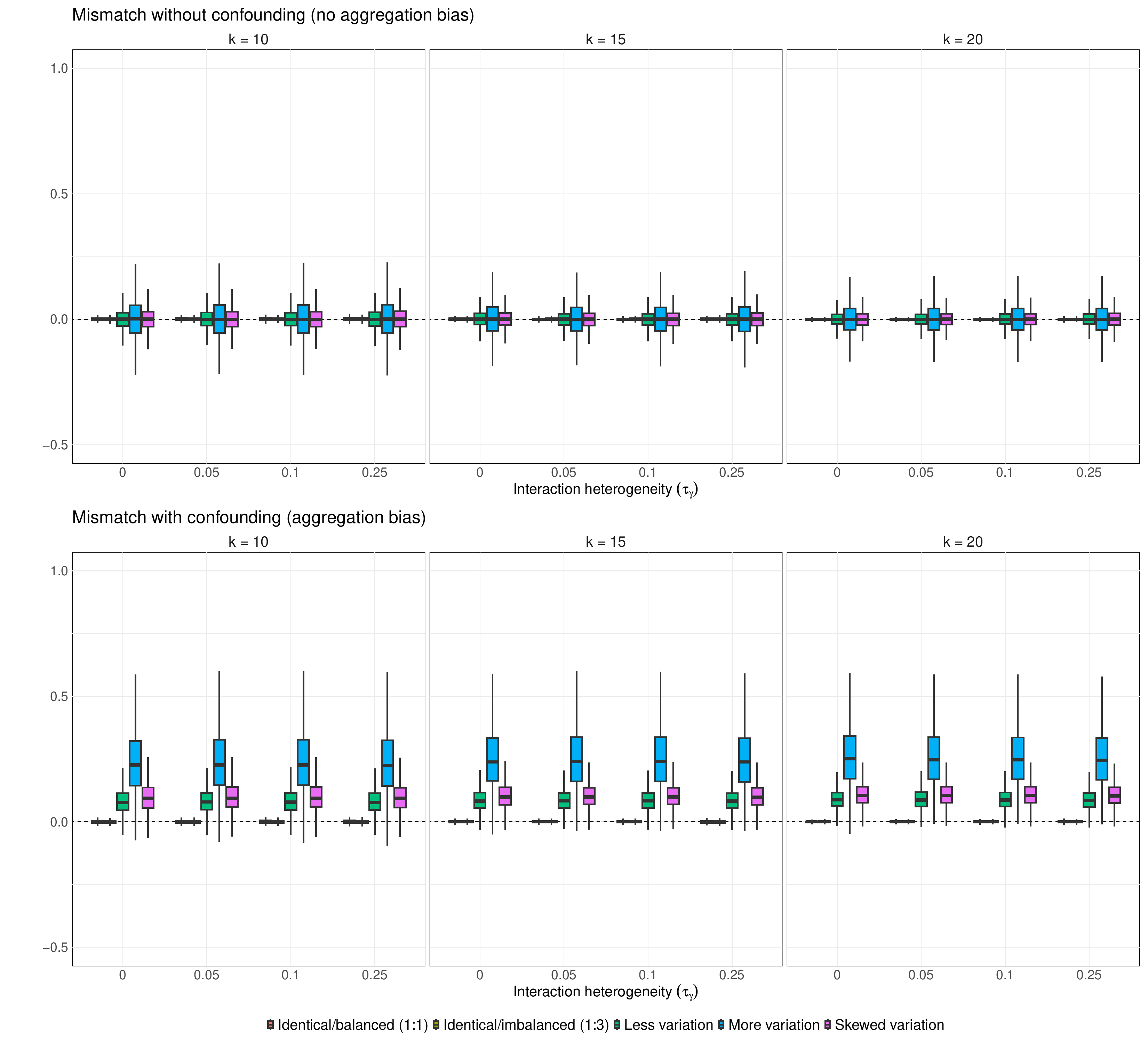}
\caption{Mismatch between subgroup-specific effect difference ($\widehat{\gamma}_{\mathrm{DA}}$) and interaction estimates ($\widehat{\gamma}_{\mathrm{AD}}$) across varying levels of interaction heterogeneity ($\tau_\gamma$).
\emph{Top vs.~bottom panels:} Bottom panels display results when aggregation bias is present (i.e., non-zero $\delta = 3$), while top panels represent scenarios without aggregation bias ($\delta=0$).
The $x$-axis represents the amount of interaction heterogeneity ($\tau_\gamma$), and the $y$-axis shows the observed mismatch $\widehat\Delta$ in \eqref{eqn:delta} for inverse-variance weights meta-analyses under the random-effects assumption.
\emph{Left to right:} Panels show increasing numbers of studies ($k = 10, 15, 20$).
Colors indicate different subgroup composition scenarios described in Section \ref{sec:DataGeneration}.
For visual clarity and file size considerations, extreme outliers are omitted from this figure}
\label{fig:DeltaMismatch_Analysis_full}
\end{figure}

Figure \ref{fig:DeltaMismatch_Analysis_full} illustrates the mismatch between the true subgroup-specific effect difference \((\widehat{\gamma}_{\mathrm{DA}}\)) and the interaction estimator \((\widehat{\gamma}_{\mathrm{AD}}\)) across varying levels of interaction heterogeneity (\(\tau_\gamma\)), numbers of studies (\(k = 10, 15, 20\)), and prevalence balance scenarios. The bottom panels show results when aggregation bias is present (\(\delta > 0\)), whereas the top panels assume no aggregation bias (\(\delta = 0\)). Colours denote different patterns of subgroup prevalence (identical/balanced, identical/imbalanced, less variation, more variation, and skewed variation).

\subsection{Extending Riley \emph{et~al}'s confounded IPD model to aggregate data}
\label{app:TheIPDprevalenceadjustedmodelbyRileyetal}
Riley \emph{et~al} [\cite{RileyEtAl2020}] provide an IPD approach that allows for subgroup-effect estimation while separating within- and across-trial evidence. The idea of explicitly distinguishing ``within-study'' and ``between-study'' evidence, and accommodating these by separate
parameters may be transferred from the IPD context to the case of meta-analysis of aggregated data. At the aggregated level (subgroups~$A$ and~$B$ of the $j$th study) this may be expressed as
\begin{eqnarray}
\begin{pmatrix}{y}_{\mathrm{A}j} \\ {y}_{\mathrm{B}j}\end{pmatrix} \mid \sigma_{\mathrm{A}j}, \sigma_{\mathrm{B}j}, \varphi, \gamma,  \gamma_{\mathrm{Agg}}, p_j, \Tau_j & \sim & \normaldistn\left( 
\begin{pmatrix} \varphi + \gamma_{\mathrm{Agg}}p_j+(0 -p_j)\gamma\\[1mm]
\varphi+ \gamma_{\mathrm{Agg}}p_j+ (1 -p_j) \gamma \end{pmatrix},
\; \begin{pmatrix} 
\sigma_{\mathrm{A}j}^2 & 0 \\
0 & \sigma_{\mathrm{B}j}^2 
\end{pmatrix} + \Tau_j \right), \nonumber\\
\label{eqn:marginal_model}
\end{eqnarray}
where $\sigma_{\mathrm{A}j}^2$ and $\sigma_{\mathrm{B}j}^2$ denote the sampling variances in the respective subgroups.
The parameters~$\gamma_{\mathrm{Agg}} := \delta+\gamma $ and~$\gamma$ reflect the two interaction estimates: the first based on commingled ``between-'' and ``within-study'' evidence and the second purely on ``within-study'' evidence. Ideally, these would match, and any discrepancy between these two estimates ($\widehat\delta=(\widehat\gamma_{\mathrm{Agg}} - \widehat\gamma) \neq 0$) would then warn meta-analysts about the so-called \emph{aggregation bias} due to confounding coming from the \emph{assumed} effect-prevalence relationship.
The accompanying between-study heterogeneity is  
\begin{eqnarray}
\Tau_j = \Tau_j^{(\tau)}  + \Tau_j^{(\tau_\gamma)}
 = \begin{pmatrix}
\tau^2 & \tau^2 \\
\tau^2 & \tau^2  
\end{pmatrix} + \begin{pmatrix}
p_j^2\,\tau_\gamma^2 & - p_j(1-p_j)\,\tau_\gamma^2 \\
 - p_j(1-p_j)\,\tau_\gamma^2 & (1-p_j)^2\,\tau_\gamma^2 
\end{pmatrix}.\label{eqn:heterogeneity}
\end{eqnarray}
where \(\tau\) is the between-trial treatment effect heterogeneity and \(\tau_\gamma\) is the interaction effect heterogeneity. 

\subsection{Complementary simulation results}

\begin{figure}[!h]
    \centering      
    \includegraphics[width=0.9\linewidth]{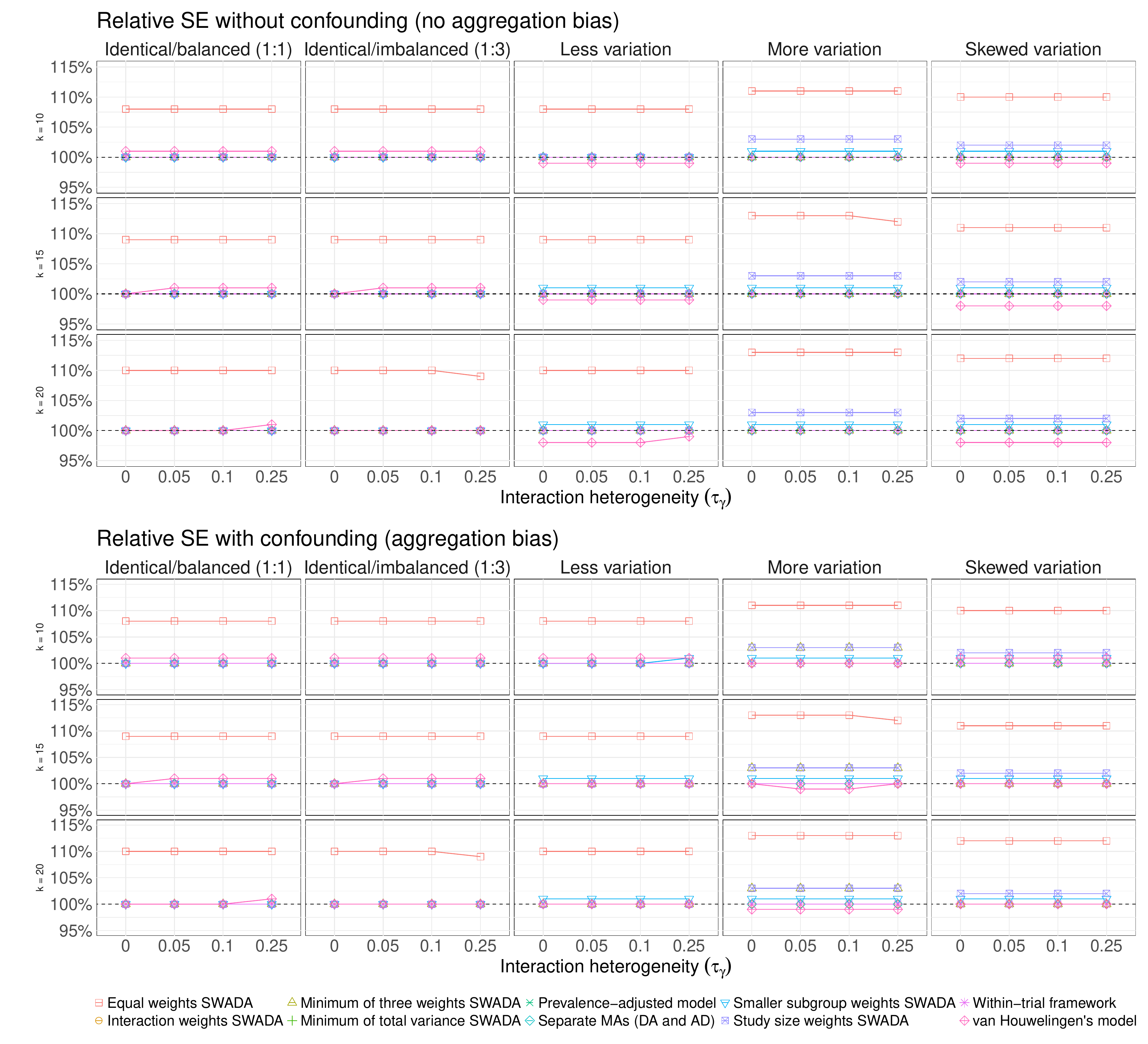}
    \caption{Ratio between interval widths for the treatment-by-subgroup interaction estimator under various methodological approaches with AD as the reference. The top panels show performance on data generated with no aggregation bias and the bottom panels under an aggregation bias assumption. The dashed horizontal line indicates the reference ratio of 100\% width (i.e. AD width), and shaded areas represent Monte Carlo confidence intervals. Each column corresponds to a different subgroup-prevalence (subgroup-composition) scenario (Section \ref{sec:DataGeneration}): identical/balanced (1:1); \(p_j=0.50\), identical/imbalanced (1:3); \(p_j=0.25\), little variation \(p_j\sim \mathrm{Uniform}(0.3,0.7)\), more variation \(p_j\sim \mathrm{Uniform}(0.1,0.9)\), and skewed variation \(p_j\sim \mathrm{triangular}(0.1,0.5)\), mode 0.1)}\label{fig:Simulation_InteractionEffectsWidth}
\end{figure}

\begin{figure}[!h]
   \centering         
   \includegraphics[width=0.9\linewidth]{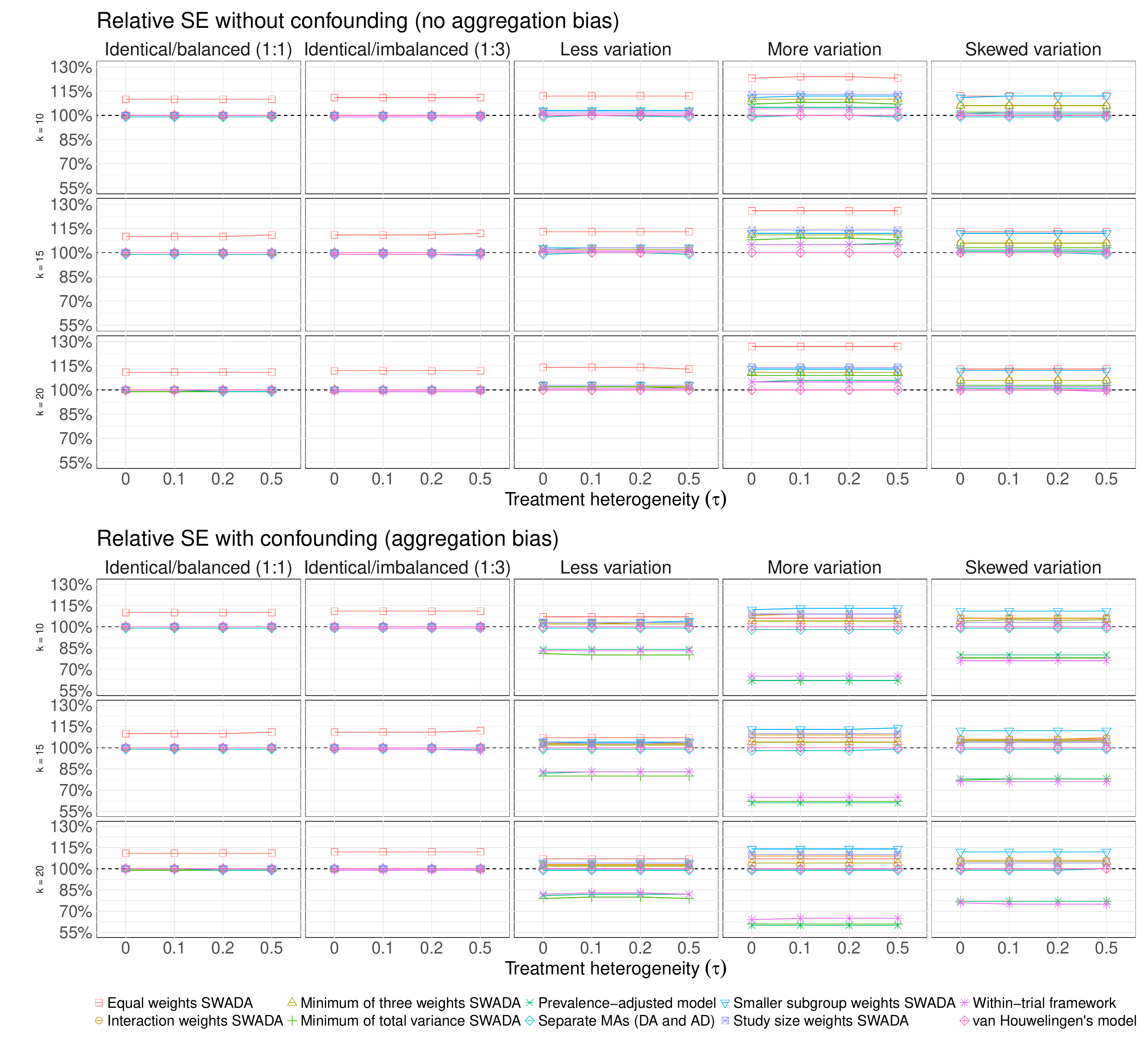}
    \caption{Ratio between interval widths for the expected reference subgroup's treatment effect under various methodological approaches with DA as the reference. Here, the reference-subgroup-specific effect refers to $\beta_A$  in Section~\ref{sec:aims}      (treatment effect at the average covariate value \(p_j = p\)), not the covariate-averaged subgroup-specific effects. The top panels show performance on data generated with no aggregation bias and the bottom panels under an aggregation bias assumption. The dashed horizontal line indicates the reference ratio of 100\% width (DA), and shaded areas represent Monte Carlo confidence intervals. Each column corresponds to a different subgroup-prevalence (subgroup-composition) scenario (Section \ref{sec:DataGeneration}): identical/balanced (1:1); \(p_j=0.50\), identical/imbalanced (1:3); \(p_j=0.25\), little variation \(p_j\sim \mathrm{Uniform}(0.3,0.7)\), more variation \(p_j\sim \mathrm{Uniform}(0.1,0.9)\), and skewed variation \(p_j\sim \mathrm{triangular}(0.1,0.5)\), mode 0.1)}    \label{fig:Simulation_SubgroupEffectsWidth}
\end{figure}

\begin{figure}[!h]
    \centering      
    \includegraphics[width=0.9\linewidth]{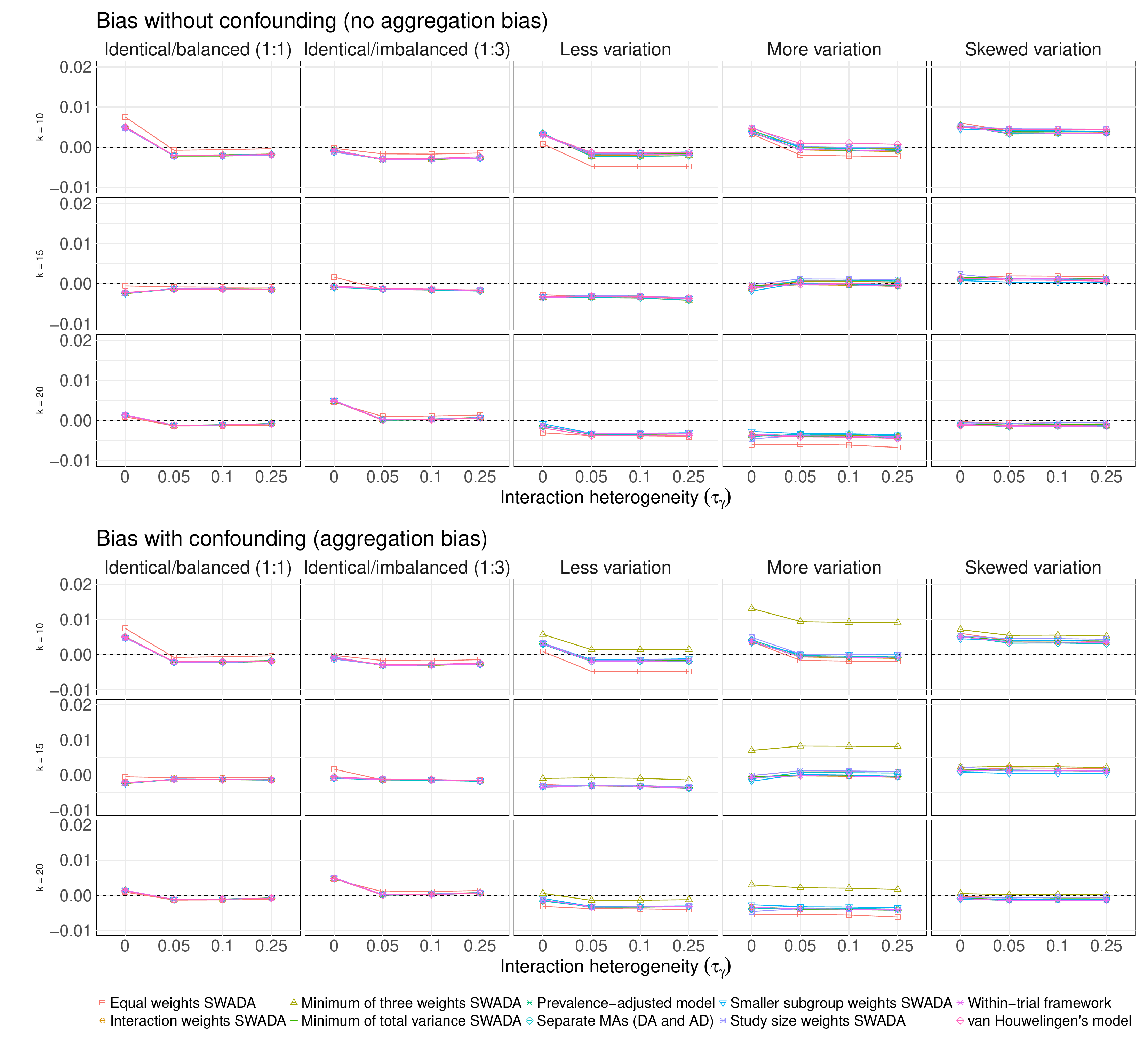}
    \caption{Bias for the treatment-by-subgroup interaction estimator under various methodological approaches. The top panels show performance on data generated with no aggregation bias and the bottom panels under an aggregation bias assumption. The dashed horizontal line indicates no bias, that is 0. Each column corresponds to a different subgroup-prevalence (subgroup-composition) scenario (Section \ref{sec:DataGeneration}): identical/balanced (1:1); \(p_j=0.50\), identical/imbalanced (1:3); \(p_j=0.25\), little variation \(p_j\sim \mathrm{Uniform}(0.3,0.7)\), more variation \(p_j\sim \mathrm{Uniform}(0.1,0.9)\), and skewed variation \(p_j\sim \mathrm{triangular}(0.1,0.5)\), mode 0.1)}\label{fig:Simulation_InteractionEffectsBias}
\end{figure}

\begin{figure}[!h]
   \centering         
   \includegraphics[width=0.9\linewidth]{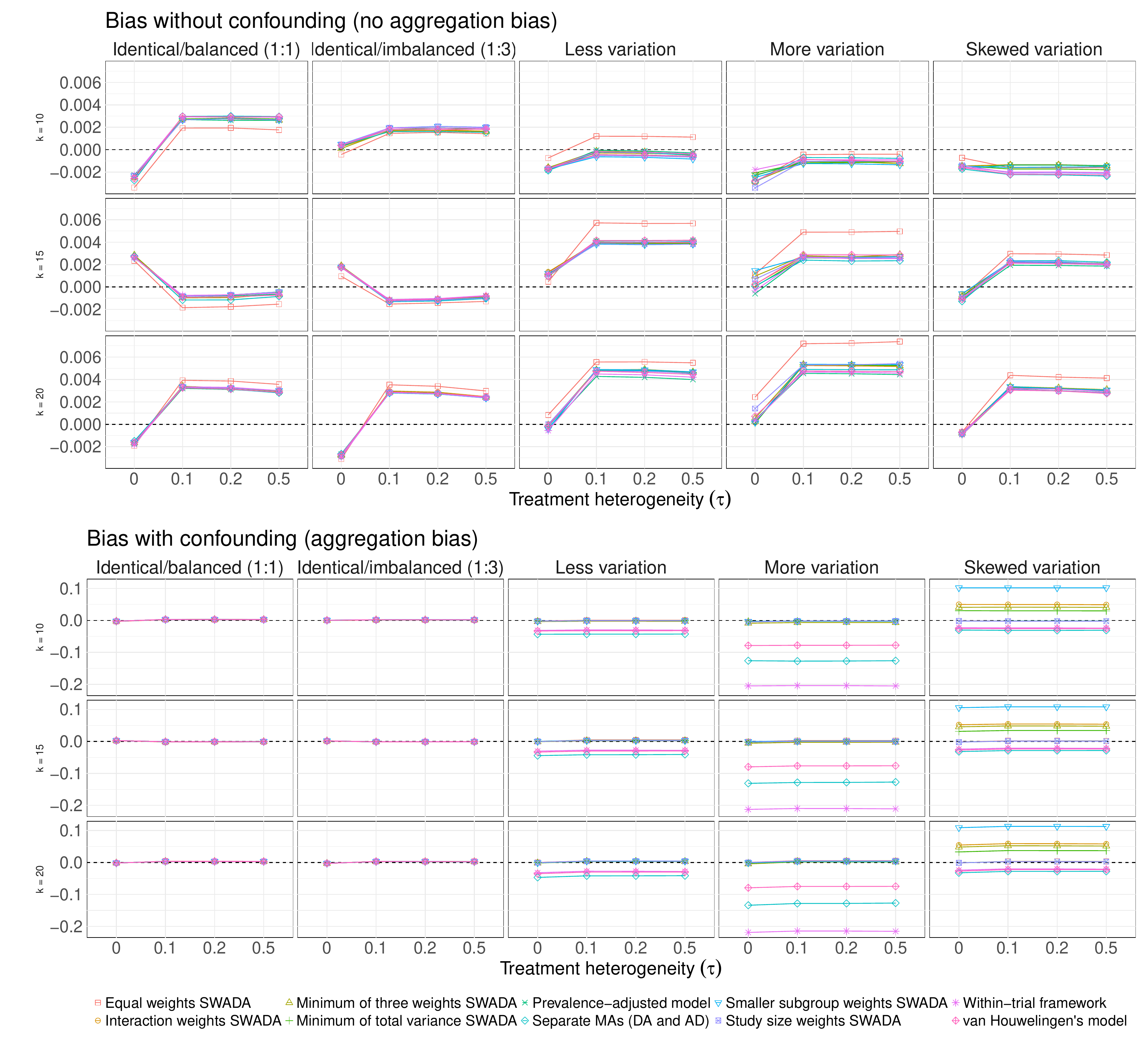}
    \caption{Bias for the expected reference subgroup's treatment effect under various methodological approaches. Here, the reference-subgroup-specific effect refers to $\beta_A$  in Section~\ref{sec:aims}      (treatment effect at the average covariate value \(p_j = p\)), not the covariate-averaged subgroup-specific effects. The top panels show performance on data generated with no aggregation bias and the bottom panels under an aggregation bias assumption. The dashed horizontal line indicates no bias, that is 0. Each column corresponds to a different subgroup-prevalence (subgroup-composition) scenario (Section \ref{sec:DataGeneration}): identical/balanced (1:1); \(p_j=0.50\), identical/imbalanced (1:3); \(p_j=0.25\), little variation \(p_j\sim \mathrm{Uniform}(0.3,0.7)\), more variation \(p_j\sim \mathrm{Uniform}(0.1,0.9)\), and skewed variation \(p_j\sim \mathrm{triangular}(0.1,0.5)\), mode 0.1)}    \label{fig:Simulation_SubgroupEffectsBias}
\end{figure}

\begin{figure}[!h]
    \centering      
    \includegraphics[width=0.9\linewidth]{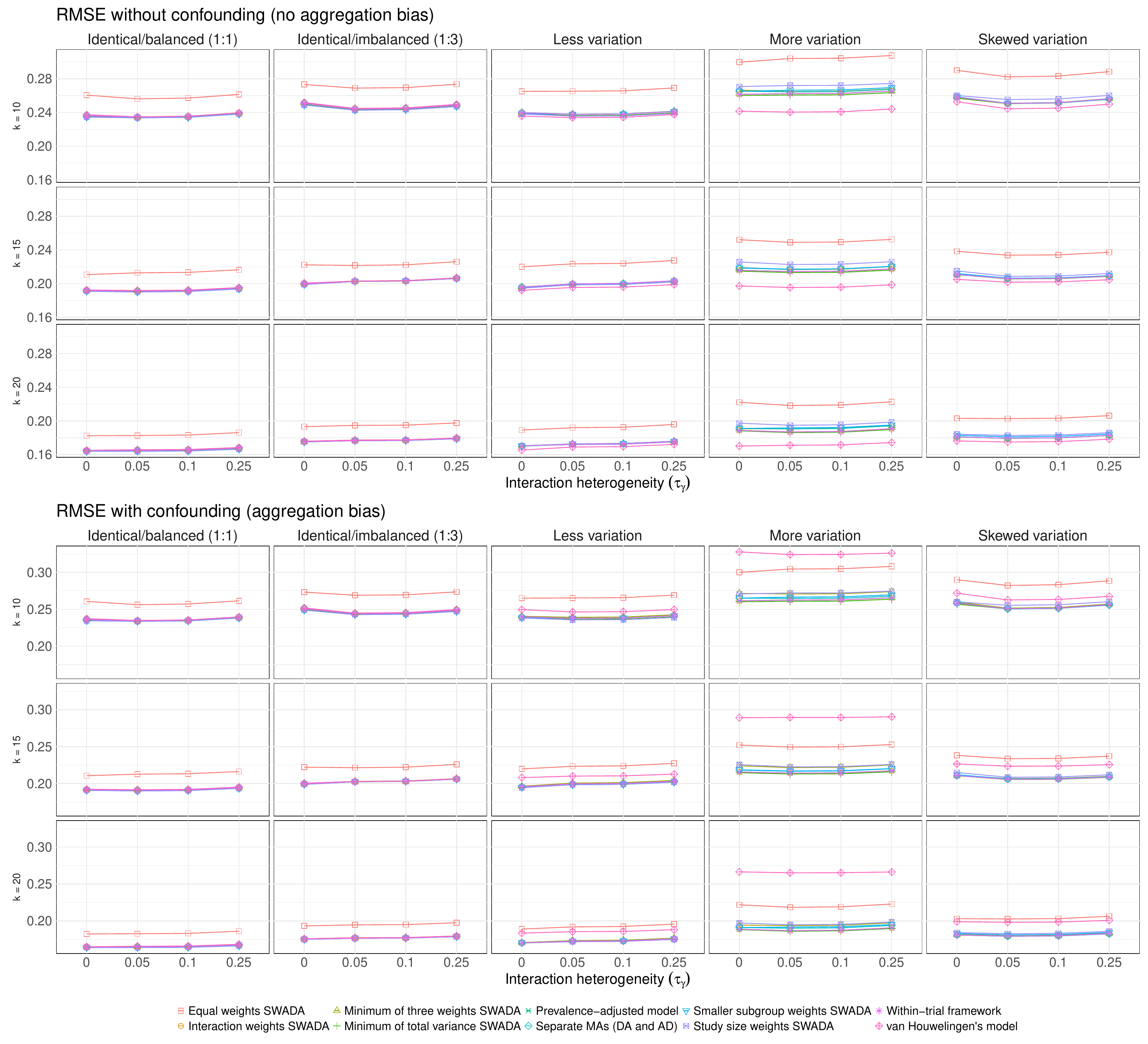}
    \caption{Root mean squared error (RMSE) for the treatment-by-subgroup interaction estimator under various methodological approaches. The top panels show performance on data generated with no aggregation bias and the bottom panels under an aggregation bias assumption. Each column corresponds to a different subgroup-prevalence (subgroup-composition) scenario (Section \ref{sec:DataGeneration}): identical/balanced (1:1); \(p_j=0.50\), identical/imbalanced (1:3); \(p_j=0.25\), little variation \(p_j\sim \mathrm{Uniform}(0.3,0.7)\), more variation \(p_j\sim \mathrm{Uniform}(0.1,0.9)\), and skewed variation \(p_j\sim \mathrm{triangular}(0.1,0.5)\), mode 0.1)}\label{fig:Simulation_InteractionEffectsRMSE}
\end{figure}

\begin{figure}[!h]
   \centering         
   \includegraphics[width=0.9\linewidth]{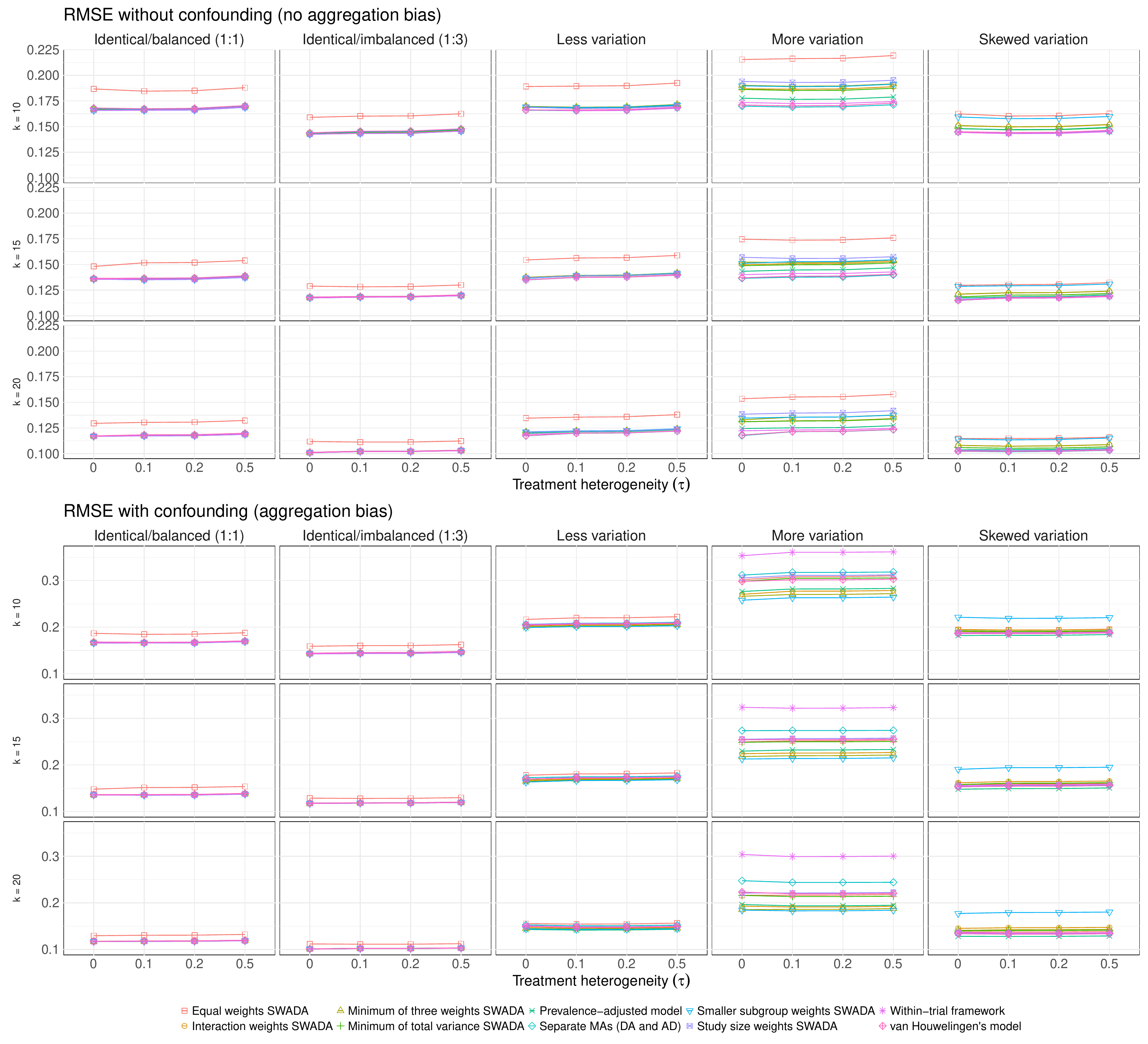}
    \caption{Root mean squared error (RMSE) for the expected reference subgroup's treatment effect under various methodological approaches. Here, the reference-subgroup-specific effect refers to $\beta_A$  in Section~\ref{sec:aims}      (treatment effect at the average covariate value \(p_j = p\)), not the covariate-averaged subgroup-specific effects. The top panels show performance on data generated with no aggregation bias and the bottom panels under an aggregation bias assumption. Each column corresponds to a different subgroup-prevalence (subgroup-composition) scenario (Section \ref{sec:DataGeneration}): identical/balanced (1:1); \(p_j=0.50\), identical/imbalanced (1:3); \(p_j=0.25\), little variation \(p_j\sim \mathrm{Uniform}(0.3,0.7)\), more variation \(p_j\sim \mathrm{Uniform}(0.1,0.9)\), and skewed variation \(p_j\sim \mathrm{triangular}(0.1,0.5)\), mode 0.1)}    \label{fig:Simulation_SubgroupEffectsRMSE}
\end{figure}

\begin{figure}[!h]
    \centering      
    \includegraphics[width=0.9\linewidth]{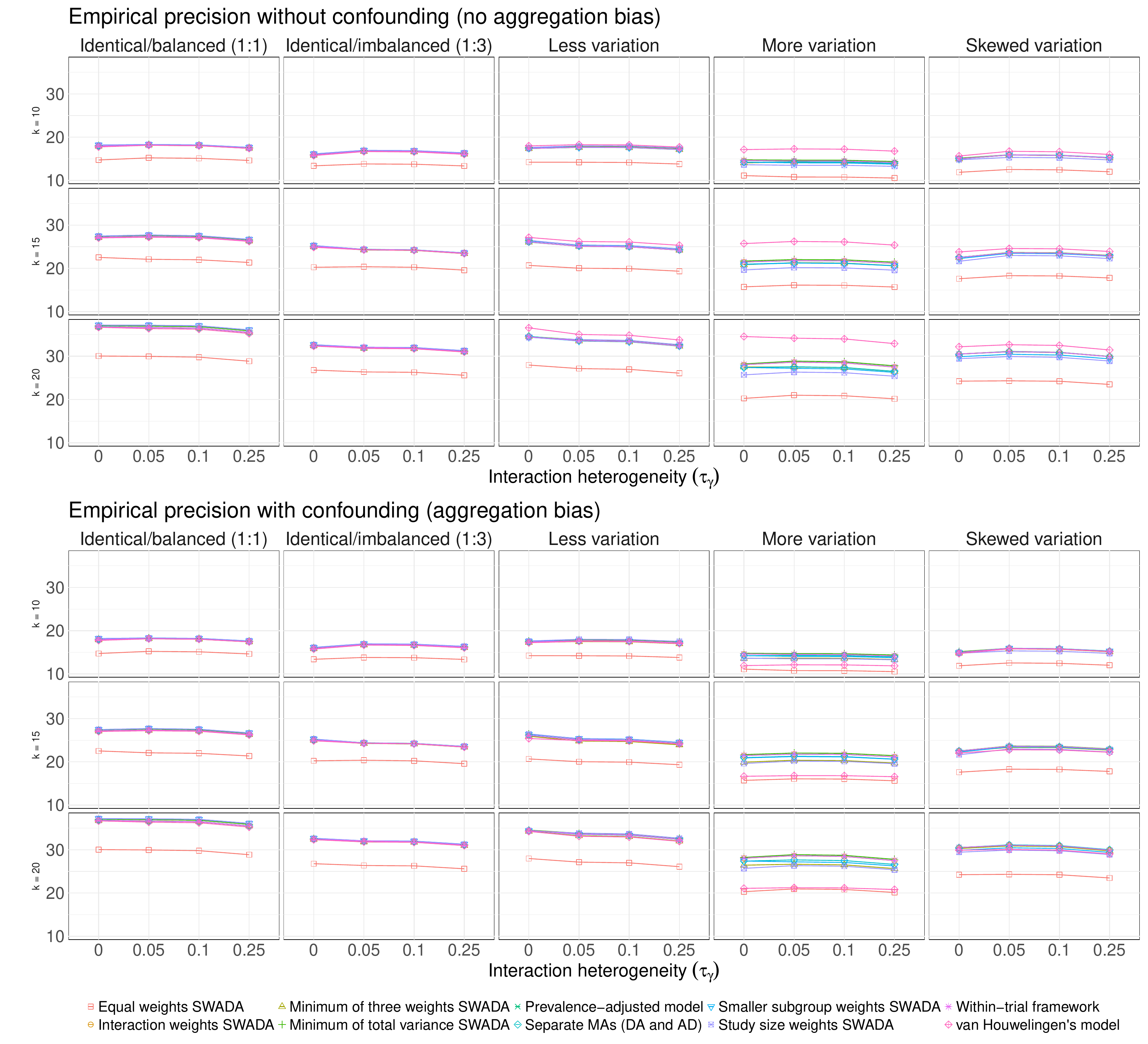}
    \caption{Empirical estimator precision (inverse empirical variance across replications) for the treatment-by-subgroup interaction estimator under various methodological approaches. The top panels show performance on data generated with no aggregation bias and the bottom panels under an aggregation bias assumption. Each column corresponds to a different subgroup-prevalence (subgroup-composition) scenario (Section \ref{sec:DataGeneration}): identical/balanced (1:1); \(p_j=0.50\), identical/imbalanced (1:3); \(p_j=0.25\), little variation \(p_j\sim \mathrm{Uniform}(0.3,0.7)\), more variation \(p_j\sim \mathrm{Uniform}(0.1,0.9)\), and skewed variation \(p_j\sim \mathrm{triangular}(0.1,0.5)\), mode 0.1)}\label{fig:Simulation_InteractionEffectsPrecision}
\end{figure}

\begin{figure}[!h]
   \centering         
   \includegraphics[width=0.9\linewidth]{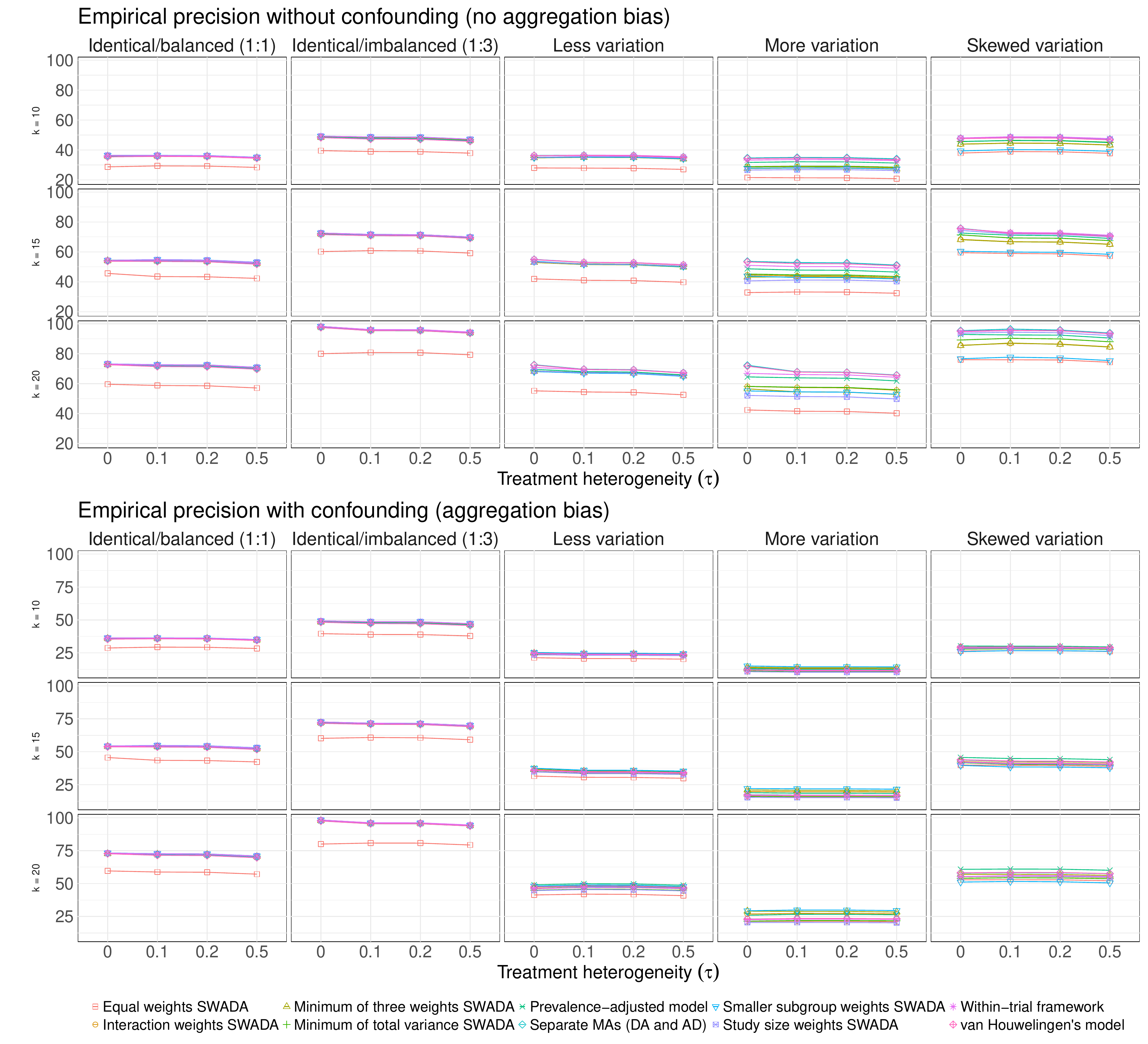}
    \caption{Empirical estimator precision (inverse empirical variance across replications) for the expected reference subgroup's treatment effect under various methodological approaches. Here, the reference-subgroup-specific effect refers to $\beta_A$  in Section~\ref{sec:aims}      (treatment effect at the average covariate value \(p_j = p\)), not the covariate-averaged subgroup-specific effects. The top panels show performance on data generated with no aggregation bias and the bottom panels under an aggregation bias assumption. Each column corresponds to a different subgroup-prevalence (subgroup-composition) scenario (Section \ref{sec:DataGeneration}): identical/balanced (1:1); \(p_j=0.50\), identical/imbalanced (1:3); \(p_j=0.25\), little variation \(p_j\sim \mathrm{Uniform}(0.3,0.7)\), more variation \(p_j\sim \mathrm{Uniform}(0.1,0.9)\), and skewed variation \(p_j\sim \mathrm{triangular}(0.1,0.5)\), mode 0.1)}    \label{fig:Simulation_SubgroupEffectsPrecision}
\end{figure}

Figure~\ref{fig:Simulation_InteractionEffectsWidth} shows the interval widths for estimating the interaction effect \((\gamma)\) across subgroup-composition scenarios, with and without aggregation bias. The top panel shows results without aggregation bias, whereas the bottom panel shows results under the aggregation-bias mechanism. Figure~\ref{fig:Simulation_SubgroupEffectsWidth} presents the corresponding interval widths for estimating the expected reference-subgroup treatment effect \((\beta_A)\). Because interval width does not necessarily reflect estimator variability when coverage differs across methods, complementary results are reported separately: bias is reported in Figures~\ref{fig:Simulation_InteractionEffectsBias} and~\ref{fig:Simulation_SubgroupEffectsBias}, RMSE in Figures~\ref{fig:Simulation_InteractionEffectsRMSE} and~\ref{fig:Simulation_SubgroupEffectsRMSE}, and empirical estimator precision in Figures~\ref{fig:Simulation_InteractionEffectsPrecision} and~\ref{fig:Simulation_SubgroupEffectsPrecision}.

\end{appendix}
\clearpage


\begin{Backmatter}


\paragraph{Funding Statement}
  Support from the \emph{Volkswagen Stiftung} is gratefully acknowledged (project ``Bayesian and nonparametric statistics - Teaming up two opposing theories for the benefit of prognostic studies in COVID-19'').

\paragraph{Competing Interests}
  The authors have declared no conflict of interest.

\paragraph{Data Availability Statement}
The datasets used in the two motivating examples
(\texttt{il6.csv} and \texttt{steroids.csv}),
together with fully reproducible \texttt{R} routines
for SWADA and for the within-trial framework,
are publicly available on Zenodo at
\url{https://doi.org/10.5281/zenodo.22812083},
with the development version available at
\url{https://github.com/rvpanaro/SWADA}.


\paragraph{Author Contributions}
Conceptualization: R.P; C.R.;T.F. Methodology: R.P; C.R.;T.F. 
Writing original draft: R.P. All authors approved the final submitted draft.



\printbibliography

\end{Backmatter}
\end{document}